\documentclass[twocolumn]{aa}  
\usepackage{graphicx}
\usepackage{txfonts}
\usepackage[usenames]{color}
\usepackage{aas_macros}
\usepackage{textcomp}
\usepackage{hyperref}
\usepackage{natbib, twoopt}
\usepackage[dvipsnames]{xcolor}
\usepackage{wasysym}
\usepackage{here}
\usepackage{float}
\usepackage{multirow}
\usepackage{lipsum,capt-of,graphicx}
\usepackage{footnote}
\makesavenoteenv{tabular}

\hypersetup{colorlinks=true,linkcolor=blue,citecolor=ProcessBlue,urlcolor=ProcessBlue}
\definecolor{green}{rgb}{0.,1.,0.}
\definecolor{orange}{rgb}{1,0.5,0}

\bibpunct{(}{)}{;}{a}{}{,}

\makeatletter
\newcommandtwoopt{\citeads}[3][][]{\href{http://adsabs.harvard.edu/abs/#3}%
{\def\hyper@linkstart##1##2{}%
\let\hyper@linkend\@empty\citealp[#1][#2]{#3}}}
\newcommandtwoopt{\citepads}[3][][]{\href{http://adsabs.harvard.edu/abs/#3}%
{\def\hyper@linkstart##1##2{}%
\let\hyper@linkend\@empty\citep[#1][#2]{#3}}}
\newcommandtwoopt{\citetads}[3][][]{\href{http://adsabs.harvard.edu/abs/#3}%
{\def\hyper@linkstart##1##2{}%
\let\hyper@linkend\@empty\citet[#1][#2]{#3}}}
\newcommandtwoopt{\citeyearads}[3][][]%
{\href{http://adsabs.harvard.edu/abs/#3}
{\def\hyper@linkstart##1##2{}%
\let\hyper@linkend\@empty\citeyear[#1][#2]{#3}}}
\makeatother

\usepackage{ulem}
\begin{document}

   \title{The SPHERE view of Wolf-Rayet 104\thanks{Based on observations collected at the European Organisation for Astronomical Research in the Southern Hemisphere under ESO programme IDs 097.D-0662 and 60.A-9639(A)}}

   \subtitle{Direct detection of the Pinwheel and the link with the nearby star}

   \author{A. Soulain\inst{1} 
          \and
          F. Millour\inst{1}
          \and
          B. Lopez\inst{1}
          \and
          A. Matter\inst{1}
           \and
          E. Lagadec\inst{1}
          \and
          M. Carbillet\inst{1}
          \and
          A. La Camera\inst{2}
          \and
          A. Lamberts\inst{3}
          \and 
          M. Langlois\inst{4}
          \and 
          J. Milli\inst{5}
          \and 
          H. Avenhaus\inst{6}
          \and 
          Y. Magnard\inst{7}
          \and 
          A. Roux\inst{7}
          \and 
          T. Moulin\inst{7}
          \and 
          M. Carle\inst{8}
          \and
          A. Sevin\inst{9}
          \and
          P. Martinez\inst{1}
          \and
          L. Abe\inst{1}
          \and
          J. Ramos\inst{10}          
          }
          
   \institute{Universit\'e C\^ote d’Azur, Observatoire de la C\^ote d’Azur, CNRS, Laboratoire Lagrange, Bd de l'Observatoire, CS 34229, 06304 Nice cedex 4, France;\\
              \email{anthony.soulain@oca.eu}
   	     \and
    Dipartimento di Informatica, Bioingegneria, Robotica e Ingegneria dei Sistemi (DIBRIS), Università di Genova, Via Dodecaneso 35, 16145 Genova, Italy\footnote{Now with Teiga srls, Viale Brigate Partigiane 16, 16129 Genova, Italy}
   		 \and
   Theoretical Astrophysics, California Institute of Technology, Pasadena, CA 91125, USA
   \and
   CRAL, UMR 5574, CNRS, Université de Lyon, Ecole Normale Supérieure de Lyon, F-69364 Lyon Cedex 07, France
   \and 
   European Southern Observatory, Alonso de Cordova 3107, Casilla 19001 Vitacura, Santiago 19, Chile
   \and 
   Max Planck Institute for Astronomy, Königstuhl 17, D-69117 Heidelberg, Germany 
   \and 
   Univ. Grenoble Alpes, CNRS, IPAG, F-38000 Grenoble, France.
   \and
   Aix Marseille Universit\'e, CNRS, LAM (Laboratoire d'Astrophysique de Marseille) UMR 7326, 13388 Marseille, France
   \and 
   LESIA, Observatoire de Paris, PSL Research University, CNRS, Sorbonne Universit\'es, UPMC, Univ. Paris 06, Univ. Paris Diderot, Sorbonne Paris Cit\'e, 5 place Jules Janssen, 92195 Meudon, France
   \and 
    Max Planck Institute for Astronomy, K\"onigstuhl 17, D-69117 Heidelberg, Germany
   }
   \date{}

 
  \abstract
   {WR104 is an emblematic dusty Wolf-Rayet star and the prototypical member of a sub-group hosting spirals that are mainly observable with high-angular resolution techniques. Previous aperture masking observations showed that WR104 is likely an interacting binary star at the end of its life. However, several aspects of the system are still unknown. This includes the opening angle of the spiral, the dust formation locus, and the link between the central binary star and a candidate companion star detected with the Hubble Space Telescope (HST) at 1''.}
   {Our aim was to  directly image  the dusty spiral or ``pinwheel'' structure around WR104 for the first time and determine its physical properties at large spatial scales. We also wanted to address the characteristics of the candidate companion detected by the HST.}
   {For this purpose, we used SPHERE and VISIR at the Very Large Telescope to respectively image the system in the near- and mid-infrared. Both instruments furnished an excellent view of the system at the highest angular resolution a single, ground-based telescope can provide. Based on these direct images, we then used analytical and radiative transfer models to determine several physical properties of the system.}
   {Employing a different technique than previously used, our new images have allowed us to confirm the presence of the dust pinwheel around the central star. We have also detected up to 5 revolutions of the spiral pattern of WR104 in the K-band for the first time. The circumstellar dust extends up to 2 arcsec from the central binary star in the N-band, corresponding to the past 20 years of mass loss. Moreover, we found no clear evidence of a shadow of the first spiral coil onto the subsequent ones, which likely points to a dusty environment less massive than inferred in previous studies. We have also confirmed that the stellar  candidate companion previously detected by the HST is gravitationally bound to WR104 and herein provide information about its nature and orbital elements.}
   {}


   \maketitle
%




\section{Introduction}\label{sec:intro}

The study of massive stars is important for many aspects of stellar evolution and cosmic enrichment. For example, a significant fraction of massive stars is part of in binary systems \citepads{2013A&A...550A.107S}. The stellar components of these binaries are good candidates to generate gravitational waves by the merging of the resulting black holes or neutron stars. 

Wolf-Rayet stars are hot stars with broad emission lines. These lines originate from a strong and optically thick wind, which is driven by radiation pressure and reaches a velocity of up to a thousand kilometers per second. This supersonic wind of WR stars exerts a major influence on the immediate surroundings of the star.

WR stars come in 3 groups: 1.) cWR -- classical, He-burning, H-poor remnants of massive stars; 2.) WNLh -- very luminous and massive main sequence late (cooler) stars with a hydrogen-enriched wind and a spectrum dominated by N III-V and He I-II lines; and 3.) [WR] \footnote{the square brackets are used to destinguish these from their massive counterparts} -- the nuclei of some planetary nebul\ae, i.e.\ low-medium mass stars on their way to becoming white dwarfs (WDs).
In this study, we focus on cWR and the final stage of massive stellar evolution.

Massive WR stars exist at the limit of exploding as a supernova of type SNIbc and are good candidates for long gamma-ray bursts \citepads{2017A&A...603A..51D}. WR stars also play a dominant role in the enrichment of the local ISM with their important mass-loss and their supernovae explosions. They are particularly one of the contributors of key chemical elements for planet formation (such as $^{14}$C or $^{26}$Al; \citeads{2010ApJ...714L..26T}).

WR stars are classified according to their emission spectra, which depend on where they fall on the WR evolutionary stage. Those of class WN exhibit many emission lines of nitrogen and helium, while class WC stars manifest emission lines of carbon, oxygen, and helium. The standard accepted sequence of evolution places WC stars at the very end of the WR stage, before the very short oxygen-rich WO stage and the final supernova \citepads{2014A&A...564A..30G}. The chronological evolution of WC stars is then classified into sub-stages from WC4 (youngest) to WC9 (oldest).

In addition, several WC8 and the majority of WC9 stars exhibit a strong infrared excess, which is reminiscent of hot and warm dust produced by the central source \citepads{1972A&A....20..333A}. These dust-producing WRs contribute to the enrichment of the interstellar medium and have extremely high dust-formation rates, with values up to \.{M} $ \rm{= 10^{-6}M_{\odot}/yr}$ \citepads{2004MNRAS.350..565H} corresponding to approximately $2\%$ of the standard accepted wind mass-loss of late WC stars \citepads{2008A&ARv..16..209P}. 


Several WR stars in binary systems with hot OB-type stars rich in oxygen and hydrogen are believed to harbour spiral, dusty structures called ``pinwheel nebul\ae'' (e.g.\ WR98a \citepads{1999ApJ...525L..97M}, WR104 \citepads{1999Natur.398..487T}, WR118 \citepads{2009A&A...506L..49M}, WR48a, WR112, WR137, WR140 (\citeads{2007ASPC..367..213M}; \citeads{2017ApJ...835L..31L}), and the Quintuplet Cluster near the Galactic Center \citepads{2006Sci...313..935T}).
These dusty spirals in such massive binary systems are interpreted as the consequence of a collision between the WR wind with that of the hot, massive companion star, which can generate a violent shock interaction at the interface of the two stellar winds (\citeads{2009MNRAS.396.1743P}, \citeads{2012A&A...546A..60L}). This wind collision zone (WCZ) between the WR and the OB components offers ideal conditions to reach critical densities and enable dust nucleation, especially where the mixing between the carbon-rich wind of the WR star and H-rich wind of the OB star becomes significant \citepads{2016MNRAS.460.3975H}. This newly-produced dust created within the shocked region then follows a ballistic trajectory forming a spiral pattern described in detail in the literature (\citeads{1999ApJ...525L..97M}, \citeads{2006Sci...313..935T}, \citeads{2008ApJ...675..698T}, \citeads{2009A&A...506L..49M}, \citeads{2017ApJ...835L..31L}). This pattern follows an Archimedian spiral in the case of a circular orbit (as for WR104), or is more complex (producing arcs) when the inner binary orbit is elliptical (as can be seen e.g.\ in WR140 or WR48a).

Among permanent dust-makers, WR104 stands out as a nearly face-on (inclination angle $i\leq16^{\circ}$)  pinwheel nebula that was detected with the aperture-masking technique on the Keck telescope \citepads{1999Natur.398..487T}. After its discovery, the system was further investigated using the same aperture-masking technique, showing the rotation of the spiral and providing some details about the spiral properties \citepads{2008ApJ...675..698T}. A radiative transfer model was developed by \citetads{2004MNRAS.350..565H} to explain the apparent flux saturation of the inner region of the spiral, detected by \citetads{1999Natur.398..487T}. This saturation was interpreted as the presence of optically thick dust in the inner region, and a dust formation region located at about 10\,mas from the central binary star.



After twenty years of observations based on reconstructed images from the 
aperture-masking technique and Fourier plane sampling, we present the first direct images of the WR104 system obtained with the Spectro-Polarimetric High-contrast Exoplanet Research (SPHERE) instrument at the Very Large Telescope (VLT) in Chile. The advantage of direct imaging is the reliability of the object's brightness distribution. 

Thanks to the new SPHERE images with high dynamic range, our aims are to: 1) constrain and confirm the general characteristics of the system (spiral step -- or radial separation between successive spiral coils, orientation, etc.), 2) provide an unprecedented view and physical description of the large scale of the pinwheel, up to 30 revolutions. Having access to these large spatial scales is essential to be able to define a set of physical models describing the dust distribution along the spiral arms and thus constrain the history of dust production over the last decade.

Our paper in organised as follows. In Section 2, we describe the SPHERE and VLT Imager and Spectrometer for mid-InfraRed (VISIR) observations and the related data reduction processes. Section 3 presents a first direct analysis of the images, followed in Section 4 by an analytical model of the dust emission in the spiral, complemented by a preliminary qualitative radiative transfer model. In Section 5, we interpret and discuss our results and finally, in Section 6, we conclude and put this work in perspective.



\section{Observations and data processing}\label{sec:obs}


\subsection{SPHERE Observations}

SPHERE is a high-performance imaging instrument equipped with an Extreme Adaptive Optics system (XAO) installed at the Nasmyth focus of Unit Telescope 3 of the Very Large Telescope \citepads{2008SPIE.7014E..18B}. Our observations were obtained between April and July 2016 in several filters across the J, H, and K near infrared bands with the IRDIS infrared camera of SPHERE, \citepads[Infra-Red Dual-beam Imager and Spectrograph,][]{2008SPIE.7018E..59D,2014SPIE.9147E..1RL}. As detailed in Table~\ref{tab:log_irdis}, we took images in many narrow-band filters; a more detailed log of the data is presented in Table ~\ref{tab:log_irdisfull}. Concurrent with our dual-band IRDIS observations, we collected data using the Integral Field Spectrometer (IFS) of SPHERE covering the YJ bands (0.95--1.35 $\mu m$) with a  spectral resolution of $R=50$, \citepads{2008SPIE.7014E..3EC}.

\begin{table}[htbp!]
	\caption{Log of the SPHERE/IRDIS observations of WR104 and two PSF calibrators (4Sgr and TYC 6295-803-1). SR is the estimated Strehl ratio performed by the AO system in the H band. The seeing measurements were taken by the 2016 Differential Image Motion Monitor of the ESO Paranal observatory. }
	\label{tab:log_irdis}
	\centering
	\renewcommand{\arraystretch}{1.3}
\begin{tabular}{p{1.7cm} l p{2cm} l l}
  \hline
  \hline
 \centering Star & MJD & Filter\protect\footnotemark & Seeing [''] & SR\\ \hline
 \centering WR104 & 57712.3 & CntH, FeII & 0.39 & 0.82\\
 \centering 4Sgr & 57512.3 & CntH, FeII & 0.56 & 0.92\\
        \hline
  \centering WR104 & 57524.1 & IFS+(H2, H3, H4) & 1.13 & 0.73\\
  \centering 4Sgr & 57524.1 & IFS+(H2, H3, H4) & 1.2 & 0.78\\
        \hline
  \centering WR104 & 57590.1 & HeI, CntJ, CntK2, CO & 0.45 & 0.87\\
  \centering TYC 6295-803-1 & 57590.1 & CntJ, CntK2 & 0.44 & 0.87\\
 \hline
\end{tabular}
\end{table}
\footnotetext{List of filters and their respective wavelengths. HeI: 1.085$\pm0.014$ $\mu m$; continuum J: 1.213$\pm0.017$ $\mu m$; continuum H: 1.554$\pm0.023$ $\mu m$, FeII lines: 1.644$\pm0.024$ $\mu m$; dual-band imaging (DBI) -- H2H3 filter pair: 1.593$\pm0.052$ and 1.667$\pm0.054$ $\mu m$, respectively, H3H4 filter pair: 1.667$\pm0.054$ and 1.733$\pm0.057$ $\mu m$, respectively; CntK2 lines: 2.266$\pm0.032$ $\mu m$; CO lines: 2.290$\pm0.033$ $\mu m$}

At most epochs, for all filters, we observed WR104 together with a non-resolved star (either Sgr 4, HD163955 of spectral type B9V, or TYC 6295-803-1), which we used to estimate the point spread function (PSF). We checked that these stars are unresolved by IRDIS ($\lambda/D\approx30$--35 mas, with $\lambda$ the wavelength of incoming radiation and $D$ the lens diameter, while the diameters are $\theta_{UD}\footnote{Diameter of the star assimilated to a uniform disk (UD).}=0.40\pm0.03$ mas for 4Sgr and $\theta_{UD}=0.40\pm0.03$ for TYC 6295-803-1) using the \textit{Searchcal/JMMC} catalogue \citepads{2006A&A...456..789B}.

The observations required 5\,h of telescope time and were taken without the SPHERE coronagraph. Due to the extreme brightness of WR104 in the H and K bands, an additional neutral-density filter was necessary in conjunction with most filters (with filter ND2, this yielded an attenuation of a factor 100 for the H and K bands, and with filter ND1, an attenuation of a factor of 10 for the J band).

All images were taken using the dithering technique, consisting of shifting the target on the detector between different integrations (a $4\times4$ pattern was used with a shift of 1 or 2 pixels).
A total of 288 images were recorded for WR104 (with $4\times4\times3$ images for each filter) in classical imaging mode, and 96 images in dual band-imaging (DBI mode), in which two different filters are used for two different parts of the camera.

The spectral response of the filters used for our multi-wavelength study (from $1~\mu m$ to $12.4~\mu m$) is shown in Figure~\ref{fig:filters}.

\subsubsection{Data reduction and astrometry}

We reduced the SPHERE/IRDIS data using an internally-developed Python program, consisting of standard calibrations (dark current subtraction, flat field correction) and cosmetic processes (bad-pixel correction). The bad-pixel map is generated using the dark images, and a median filter is applied to be compared with the raw dark image. For classical imaging, using the same filter, IRDIS detector frames are first split in two parts. Then, all frames are shifted, added, and averaged using dithering information and cross-correlation\footnote{We used a cross-correlation to determine the shift between the two parts of the detector in classical imaging mode.} in both parts of the camera. In DBI mode, two different filters are used in both parts of the detector and processed separately. A second cosmetic step is then applied the images to correct for the residual hot pixels.

The IFS data were reduced using the SPHERE Data Reduction and Handling
(DRH) automated pipeline \citepads{2008SPIE.7019E..39P} at the SPHERE
Data Center\footnote{\url{http://sphere.osug.fr}} (DC, \citeads{2017arXiv171206948D}). We applied basic corrections
for bad pixels, dark current, and flat field and complemented the DRH pipeline with additional
steps to improve the wavelength calibration, 
cross-talk, and bad pixel correction \citepads{2015A&A...576A.121M}.

Absolute orientation and pixel scales of the images were calculated using the parallactic angle (PARANG) and the absolute calibration provided by the SPHERE consortium. The current respective estimates of the pixel scale and true North (TN) angle for IRDIS are 12.255$\pm$0.009 mas/pixel and TN=\mbox{-1.75$\pm$0.08$^{\circ}$} \citepads{2016SPIE.9908E..34M}. Our images were obtained using the pupil-stabilized mode, and therefore they were also derotated from the zero point angle of the instrument pupil (PUPIL$_{\text{offset}}$). The measured values are stable around an average value of -135.99$\pm$0.11$^{\circ}$. For IFS, the provided calibrated values are a pixel scale of 7.46$\pm$0.02 mas/pixel and a North angle of TN=-102.18$\pm$0.13$^{\circ}$ \citepads{2016SPIE.9908E..34M}. Finally, the orientation angle $\alpha_{corr}$ was calculated for each wavelength to align North upward and East to the left, according to: 

\begin{equation}
\label{eq:orient}
\alpha_{corr} = -(\rm{PARANG} + TN + PUPIL_{\text{offset}}).
\end{equation}

The uncertainties of SPHERE data are computed taking the speckle noise,  photon noise, and detector noise into account. We determine the errors for each pixel using the unbiased standard deviation calculation on all the individual frames (almost 300 frames for each filter). We specifically follow Eq.~\ref{Eq:unbiased_STD} for random variables following a normal distribution:

\begin{equation}
\label{Eq:unbiased_STD}
S = k_n\left[\frac{1}{n-1}\sum_i(X_i-M)^2\right]^{1/2},
\end{equation}
\begin{equation}
\label{Eq:corr_STD}
k_n=\sqrt{\frac{n-1}{2}}\frac{\Gamma\left(\frac{n-1}{2}\right)}{\Gamma\left(\frac{n}{2}\right)},\;\;\;\;\;\;n\ge2,
\end{equation}
where, n is the frame number, M the mean value of the cube (the final image in this case), $X_i$ the individual pixel value, $k_n$ the corrector factor, and $\Gamma(x)$ the Gamma function (Euler integral of the second kind).

Figure \ref{fig:all_red} presents all the reduced images of WR104 and the associated calibrators. Accounting for the orbital period of the system of 241.5\, days, the images were all rotated to share the same orbital phase as that of 21 of July 2016.

\subsubsection{Image deconvolution}

SPHERE is an instrument designed to directly detect planets around stars other than our own Sun. As a consequence, the instrument is designed to provide a very high Strehl ratio ($\geq90$\% in band K; \citeads{2014A&A...572A..85Z}), meaning that deconvolution is an easier process than with previous AO-equipped instruments. We deconvolve the images of WR104 using the associated PSF and the Scaled Gradient Projection (SGP) algorithm \citepads{2009InvPr..25a5002B}, implemented in the IDL programming language within the \texttt{AIRY} image reconstruction package \citepads{2002A&A...387..733C}, developed within the \texttt{CAOS} Problem-Solving Environment \citep{carbilletlacamera2017}\footnote{See also \url{http://lagrange.oca.eu/caos}.}. The IFS data were deconvolved with a standard \mbox{Richardson--Lucy} algorithm (RL) implemented in Python \citepads{1974AJ.....79..745L}.

Considering the Poisson nature of the noise, the deconvolution problem is based either on the minimization of the \mbox{Kullback--Leibler} functional or the Csisz\'ar I-divergence \citep{csiszar1991} to which we can add a penalty term weighted by a real (positive) number, called a regularization parameter.

SGP is a gradient method that permits acceleration of the deconvolution process. RL can be seen as a particular case of the scaled gradient method, with a constant step-length. In contrast, SGP is an optimization method based on an adaptive strategy for the step-length parameter and is more efficient (from a computational point-of-view) than RL. In all cases, the algorithms are iterative and  must be stopped before the noise amplification exceeds the fit of the data. In fact, RL-based methods need an early stoppage of the iterations  to avoid the so-called checkerboard effect in the case of diffuse objects \citepads{2009InvPr..25l3006B}.

In the case of RL, to determine the best compromise between the deconvolution efficiency and artefact creation, we tested a range of numbers of iterations between 10 and 300. We found that 60 iterations avoid any artefact creation and increase of the noise. In the case of SGP, we stopped the iterations when the objective function (sum of the Kullback--Leibler function and the regularization term) was approximately constant, given a tolerance of $10^{-7}$. In addition to the main SGP deconvolution process, we also applied a pre-processing of the reduced PSF and acquired images. During pre-processing, we first set the minimum flux of each image's data to zero, then normalized the PSF to a unit integral, and added a small constant to the images  in order to avoid divisions by zero during the deconvolution process. After this, the background value was evaluated and then considered within the algorithm (see \citeads{2009InvPr..25l3006B}).


Moreover, in the case of SGP, and to avoid the above-mentioned checkerboard effect when it appears (which was always the case except for  the K band data), we used a second-order Tikhonov regularization defined by the discrete Laplacian of the unknown object. The choice of the associated regularization parameter ($\beta$) was chosen in each specific data case as a trade-off between the checkerboard effect and a regularized reconstruction of the observed spiral shape which was too smooth. The goal was to obtain the best possible fit of the spiral shape. In practice, the SGP deconvolution process of the data leads to:
\begin{itemize}
\item no regularization necessary in the K band (114 iterations to reach the fixed tolerance of $10^{-7}$) and in the CO filter (169 iterations);
\item $\beta$=0.01 (94 iterations) in the H band;
\item $\beta$=0.05 (188 iterations) in the FeII filter;
\item $\beta$=0.1 (82 iterations) in the J band.
\end{itemize}

It is clear here that shortening the wavelength (and hence roughly lowering the data Strehl ratio) implies increasing the regularization needed for the reconstruction, which is what is logically expected.

\begin{figure}[htbp!]
	\centering
		\includegraphics[width=.48\textwidth]{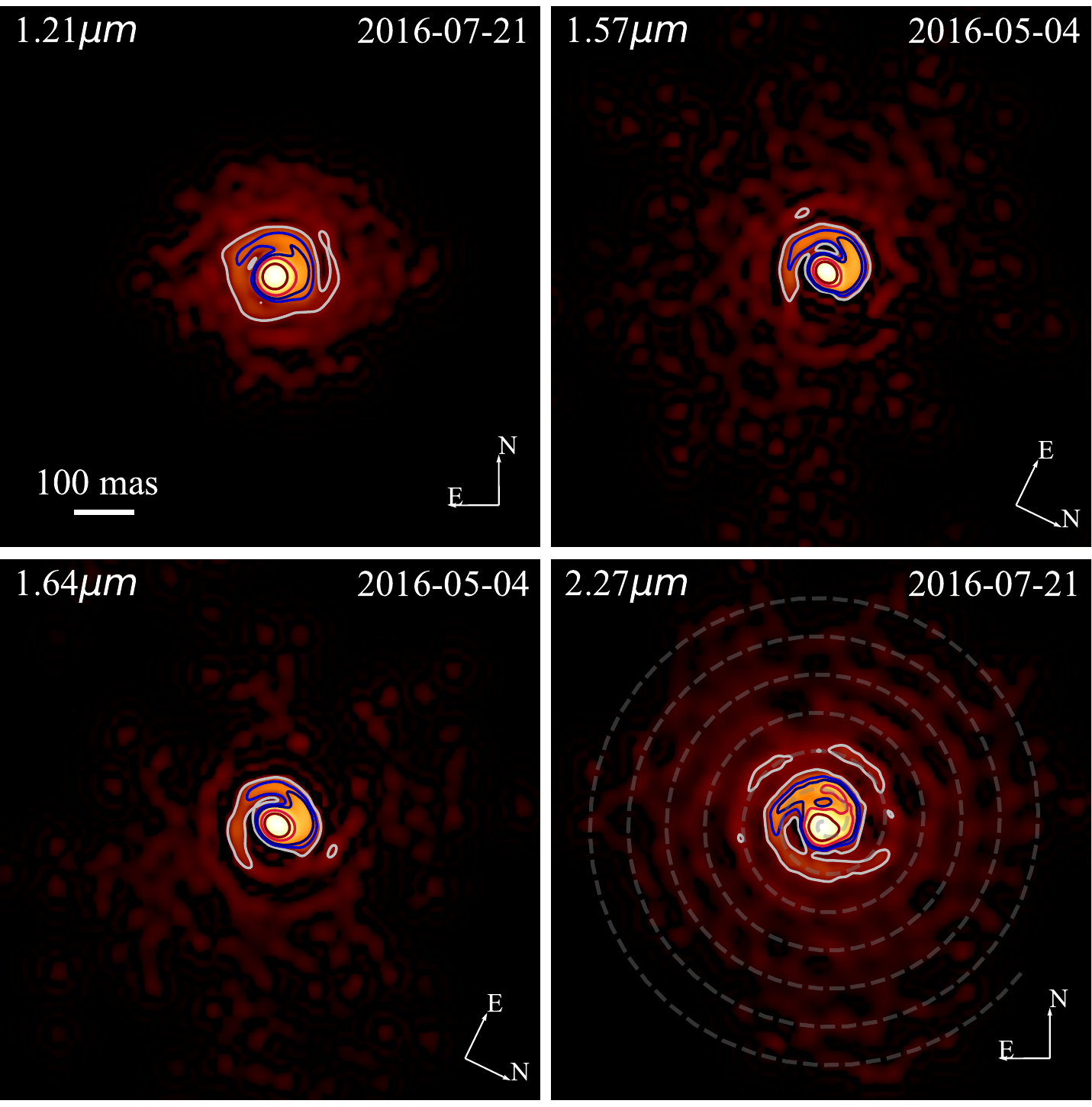} 
	\caption[example] 
	{\label{fig:deconv_carb_2017} 
	SGP deconvolution in J, H, and K bands as well as the FeII filter. Contour levels are 1, 5, 10, 20, 50\% of the peak. All images are cophased to the second epoch at 21 July 2016. The images are represented with a power normalization scale of 0.3. The field of view is 980 mas. We can clearly see the spiral pattern and the first revolution in all reconstructed images.}
\end{figure}

In Fig.\,\ref{fig:deconv_carb_2017}, we show the deconvolution of WR104 in the J, H, and K bands as well as the FeII filter. The deconvolved images exhibit a clear spiral pattern, confirming for the first time with direct imaging the pinwheel nature of WR104 (aperture masking is considered here an indirect imaging technique, since it involves interferometric techniques and image reconstruction). More specifically, in the K band deconvolved images, five turns of the spiral are detected. On the other hand, the raw images show flux in up to ten turns of the spiral (see Sect.\,\ref{curvilinear})

\vspace{0.5cm}

\subsection{VISIR observation}

VISIR is a mid-infrared imager and spectrograph installed at the Cassegrain focus of Unit Telescope 3 of the VLT \citepads{2004Msngr.117...12L}.
Our observations were obtained during 1\,h of Science Verification Time (SVT) in March 2016 with the new Raytheon AQUARIUS 1024$\times$1024 pixel arrays and the new coronagraphic mode. We used the Annular Groove Phase Mask \citepads[AGPM-$12.4\, \mu m$][]{2012SPIE.8446E..8KD} and the 4-Quadrant Phase Mask \citepads[4QPM-$10.5\, \mu m$][]{2014SPIE.9147E..0CK}. Chopping and nodding were used to remove the sky background and thermal background. The chopping was parallel to the nodding, with a chop throw of 10''. Note that with the chopping, we obtain images alternatively with and without the coronagraph centred on the target. We observed the PSF calibrator star HD169916 (spectral type K0III) unresolved by VISIR ($\lambda/D\approx300$ mas, while $\theta_{UD}=3.90\pm0.21$ mas \citepads{2005A&A...431..773R}) in the N band. The same star was also used as photometric calibrator with the corresponding absolute flux tabulated in \citetads{1999AJ....117.1864C}.

\begin{savenotes}
\begin{table}[htbp]
	\centering
	\caption{Log of the VISIR observations of WR104 and HD169916 (PSF calibrator). The seeing measurements were not communicated (twilight observation). The weather conditions were very poor for the AGPM observation (no active optics correction, high water-vapour >2.4mm).}
	\label{tab:log_visir}
	\renewcommand{\arraystretch}{1.3}
	\begin{tabular}{c c c c c}
		\hline
		\hline
		Star & MJD\footnote{MJD = Modified Julian Day} & Filter & Airmass & Seeing ['']\\
		\hline
		WR104 & 57469.40  & 10\_5\_4QP & 1.05 & nc\footnote{Not communicated by ESO at the time of observation (twilight).} \\
		WR104 & 57469.41  & 12\_4\_AGP & 1.02 & nc\\
		HD169916 & 57469.41  & 10\_5\_4QP & 1.07 & nc\\
		HD169916 & 57469.42  & 12\_4\_AGP & 1.02 & nc\\
	\end{tabular}
\end{table}
\end{savenotes}

\subsubsection{Data reduction}

The VISIR data were reduced with the official ESO-pipeline, which cleans the data from bad pixels. Nodding images
are created by averaging the images in the two positions of the
chopper. Since the images were obtained with a coronagraph with chopping in parallel mode, the final image contains a column of three images aligned along the North/South axis. The first and last images are two negative images without the coronograph, while the central image is with the coronagraph and has twice the integration time of the other two.

\begin{figure}[htbp!]
		\centering
		\includegraphics[width=.48\textwidth]{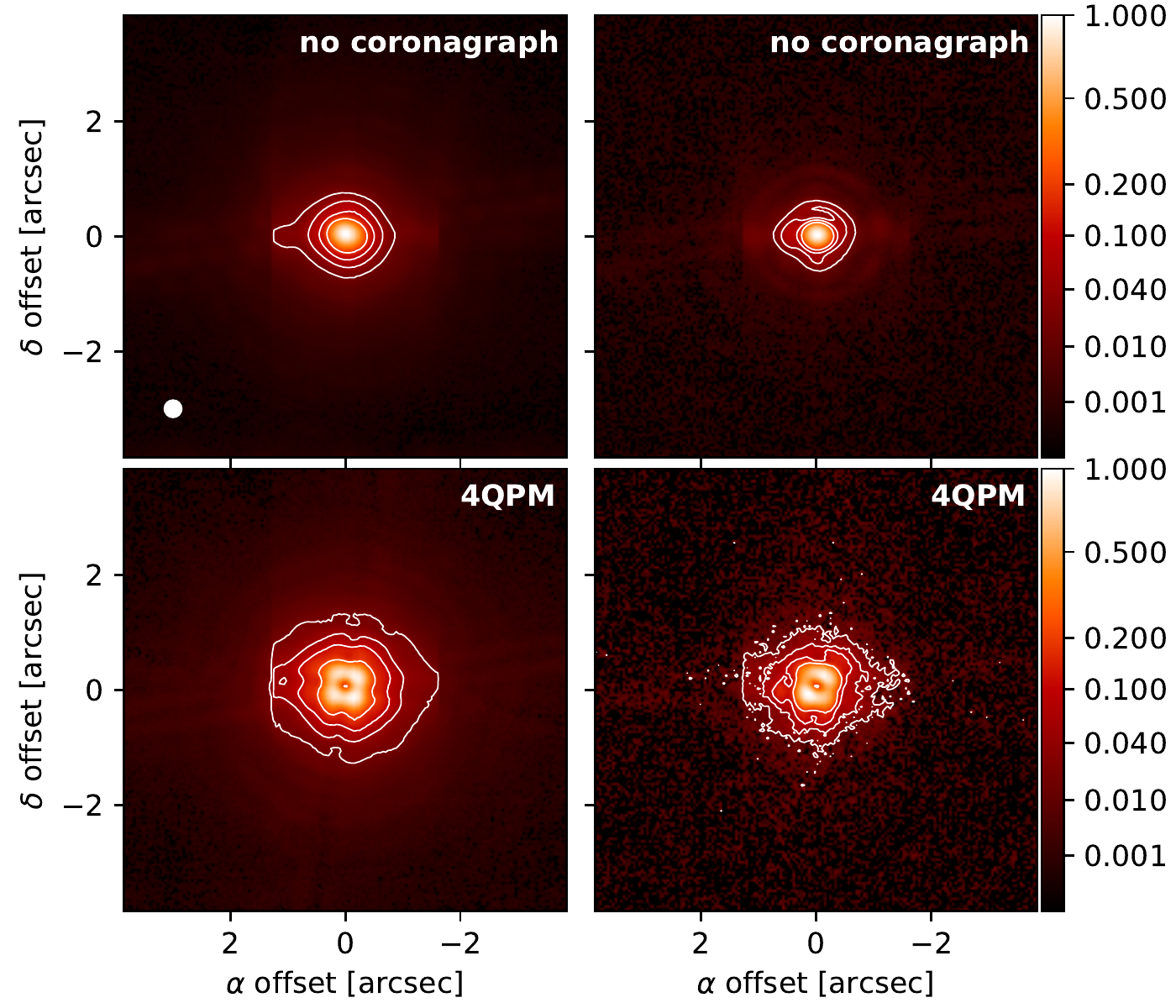}
		\caption{\label{fig:visir_QPM}QPM coronagraph at 10.5\,$\mu m$. \textbf{Left panels}: WR104, \textbf{Right panels}: PSF reference star HD169916. The top panels correspond to classical imaging, while the ones correspond to the coronograph mode. All images are displayed with a power normalization scale of 0.3. Contours at 2$\%$, 5$\%$, 10$\%$, 20$\%$ of the maximum are represented. The theoretical resolution limit by the Rayleigh criterion of $1.22\lambda/D$ is also shown as a white circle in the upper left panel.}
\end{figure}

\begin{figure}[htbp!]
		\centering
		\includegraphics[width=.48\textwidth]{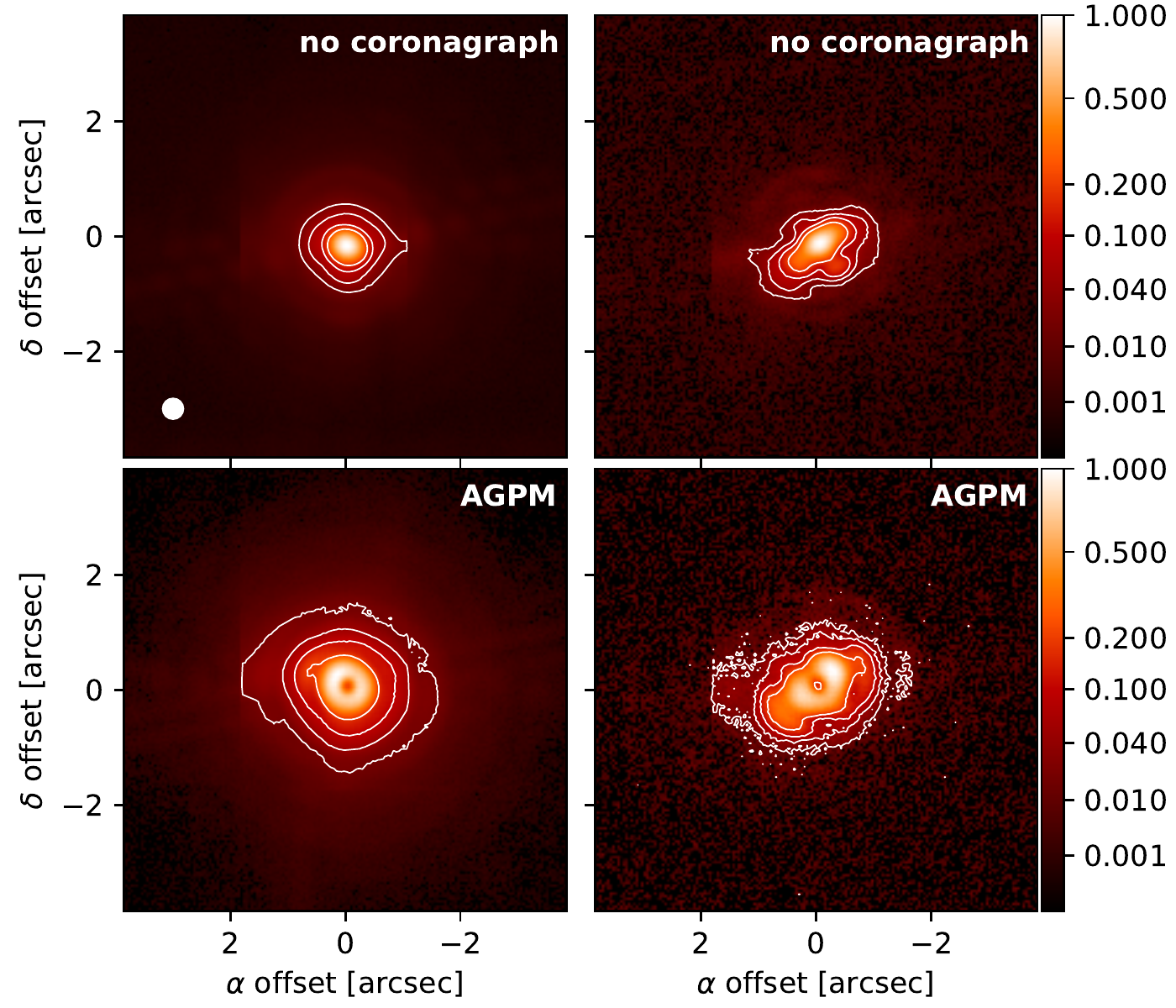}
		\caption{\label{fig:visir_AGPM}AGPM coronagraph at 12.4\,$\mu m$. Same description as for Fig.\ref{fig:visir_QPM}.}
\end{figure}

The corresponding images, in classical mode on top and coronagraphic mode at the bottom, are shown in Fig.~\ref{fig:visir_QPM} at 10.5$\,\mu m$ and  Fig.~\ref{fig:visir_AGPM} at 12.4$\,\mu m$. WR104 is clearly more extended than the PSF reference star. This is confirmed by the radial profile of Fig.~\ref{fig:radial_prf}. We cannot distinguish any relevant structure around WR104, especially at the location of its second companion star (see Section~\ref{sec:companion}). We also cannot confirm the spiral shape far from the star because of the relatively lower spatial resolution of the VISIR instrument compared to SPHERE (250\,mas vs 30--35\,mas). Nevertheless, dust is present far from the star, at least up to 2 arcsec. This implies that a fraction of the dust can survive after many orbits (30 orbits = 2''). However, we are unable to measure this fraction with the data at hand. 

We note the low quality of the PSF reference star obtained with the AGPM coronagraph. The PSF seems to be elongated in the NE-SW direction. This issue may be due to a very low wind speed, favouring dome turbulence, and/or to the fact that the observations were made at the very end of the night, during twilight, after a relatively bad night (seeing $\geq$1.2 arcsec). Therefore, the AGPM data set, especially regarding the PSF reference star, needs to be treated with caution.


\section{Data analysis}


\subsection{Radial profile}

The radiation intensity profiles at different wavelengths can tell us more about the type of physics behind the observed emission. Considering a smooth spherical shell with radius $\rm{r}$, a power-law intensity profile with a power index of -2 (i.e., $\rm{I(r)\propto r^{-2}}$) can be associated to optically thin emission, while a different power index indicates a more complex process \citepads[thermal emission, back warming, etc.][]{1986rpa..book.....R}

\begin{figure}[htbp]	
\centering			
\includegraphics[width=0.46\textwidth]{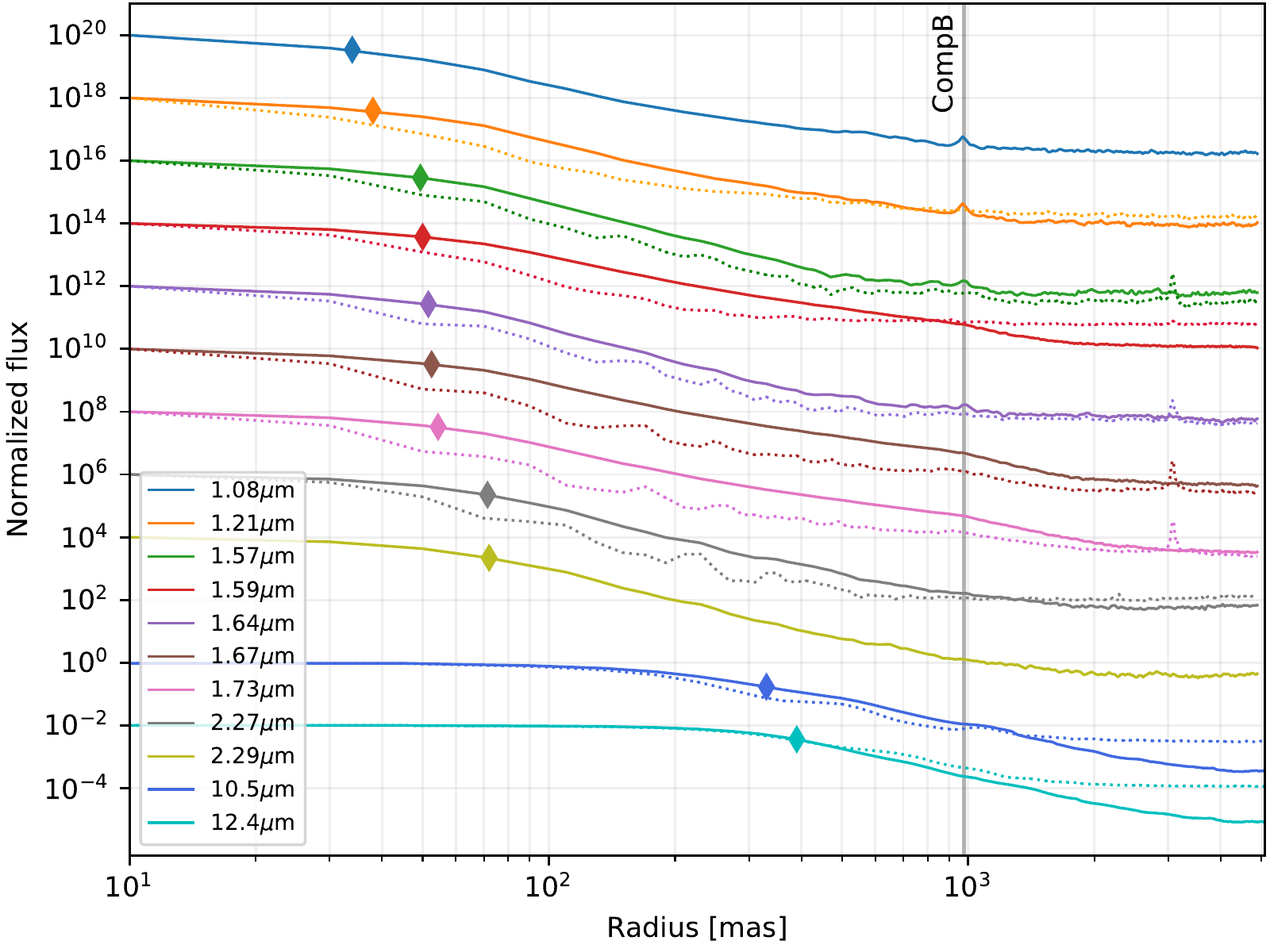}\caption{\label{fig:radial_prf} {\bf Solid lines}: Mean radial profile of the non-deconvolved SPHERE and VISIR images of WR104. All profiles are shifted for readability reasons. {\bf Dotted lines:} radial profiles of the PSF reference star. The diamonds indicate the theoretical angular resolution of the 8\,m telescope at the given wavelength. We also show  position of the companion B star of WR104 (vertical line) for information (see Section~\ref{sec:companion}).}
\end{figure}

We compute the azimuthally-averaged radial profiles on our raw (non-deconvolved) images for all available filters between 1.08 and 12.4\,$\mu$m. All profiles are normalized to 1 at the peak and shown  in Fig.~\ref{fig:radial_prf} for comparison. 

We compute the power-law index of the intensity decrement
for all filters (the q parameter in Table~\ref{tab:fit_pow}). In practice, to avoid any problems of sub-resolution, we select a zone of the profile that is unaffected by the limited resolution of each instrument at the wavelength considered (between 40 mas in the J band and 400 mas in the N band). We only consider the region above the noise floor (around 700 mas for SPHERE and 3 arcsec for VISIR) and vary the fitting region a few tens of mas around these limits to reliably estimate the uncertainties of the power-law index. The resulting indices are presented in Table \ref{tab:fit_pow}. Since the uncertainty on the radial profile varies widely across the different filters, the chi-squared values $\rm \chi^2_{red}$ are  not completely representative of goodness of fit. In Fig.\ref{fig:rad_prf_quality}, we present the results of the different fits with the associated uncertainties in the data. The figure shows the differences between the filters and the clear deviation from a power-law in the J band (CntJ and HeI filters).

\begin{table}[htbp]
	\centering
	\caption{ Power law fits of the average radial profiles. The q parameter corresponds to the power-law index fitted in the selected zone of the radial profile (see text for details). The radial profiles are presented in Fig.\ref{fig:rad_prf_quality}.}
	\label{tab:fit_pow}
	\renewcommand{\arraystretch}{1.3}
	\begin{tabular}{c c c c}
		\hline
		\hline
		Filter & Wavelength [$\mu$m] & q & $\chi_{red}^2$\\
		\hline
		HeI  & 1.08 & $-2.7\pm0.3$ & 644\\
		CntJ & 1.21 & $-3.0\pm0.2$ & 33\\
		CntH & 1.57 & $-3.3\pm0.6$ & 3\\
		H2   & 1.59 & $-2.4\pm0.2$ & 0.01\\
		FeII & 1.64 & $-3.0\pm0.6$ & 8.2\\
		H3   & 1.66 & $-2.5\pm0.2$ & 1.5 \\
		H4   & 1.73 & $-2.4\pm0.2$ & 7.7\\
        CntK2& 2.27 & $-3.0\pm0.2$  & 1.1\\
		CO   & 2.29 & $-3.0\pm0.2$ & 0.3\\
        QPM$\_$10.5 & 10.5 & $-2.7\pm0.2$ & 3.8\\
        AGPM$\_$12.4& 12.4 & $-2.8\pm0.2$ & 2.5\\
	\end{tabular}
\end{table}


\subsection{Curvilinear profile of the pinwheel}
\label{curvilinear}



\begin{figure}[htbp]	
\centering			
\includegraphics[width=0.46\textwidth]{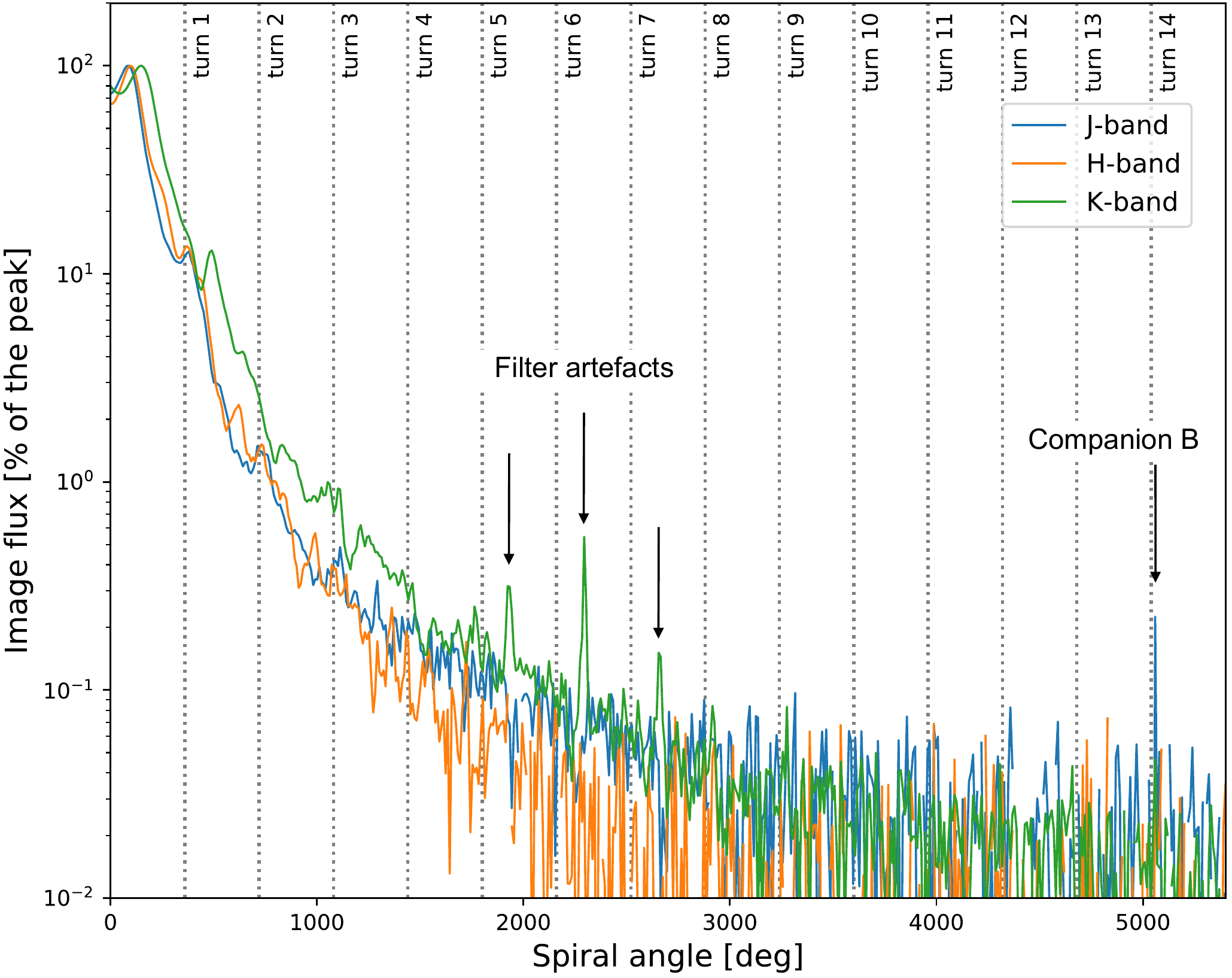}\caption{\label{fig:curviflux_red} Image flux along the fitted spiral (J, H, and K bands for the non-deconvolved images). Black dotted lines correspond to the 15 first turns of the spiral. We also note the presence of artefacts due to the K band filter.}
\end{figure}

Given the complexity of the dust distribution, we follow \citetads{2006Sci...313..935T} and compute the flux as a function of angular displacement along the spiral. With our data, we are able to detect the flux along the spiral for up to 15 coils \citepads[compared to 2 in][]{2006Sci...313..935T}, as can be seen in Fig.~\ref{fig:curviflux_red}.

First, we adjust the deconvolved images to our phenomenological model, further described in Section \ref{simplemodel}. This first fit is then refined with a comparison to the raw (non-deconvolved) image. We then use the fitted origin, orientation, and step of the spiral to compute the curvilinear flux over 15 turns in the J, H, and K bands and extract the flux at the expected position of the spiral represented by the grey, dashed line in the lower-right panel of Fig.~\ref{fig:deconv_carb_2017}. Using a cubic interpolation on the pixel coordinates, we compute the flux value 10 degree increments of the spiral's azimuthal coordinates.

The resulting flux exhibits a similar shape in all filters, with the best signal-to-noise ratio in the K band. The flux decreases monotonically with radius over the first two to three coils, without any clear breaks. The ups and downs observed in the profile (especially after the first coil) are not real and seem to come from the convolution process with the PSF. Similar to \citetads{2006Sci...313..935T}, we also observe a systematic offset between the start of the spiral and the brightest pixel. This maximum, occurring at an azimuth of 90$^\circ$ (J and H bands) and 110$^\circ$ (K band), is consistent with the offset of $90^\circ$ in azimuth (or 12$\,$mas in distance), attributed by \citet{2006Sci...313..935T} to the dust formation location locus.

\subsection{Spatially resolved spectroscopy with IFS}

\begin{figure}[htbp]
		\centering
			\includegraphics[width=0.46\textwidth]{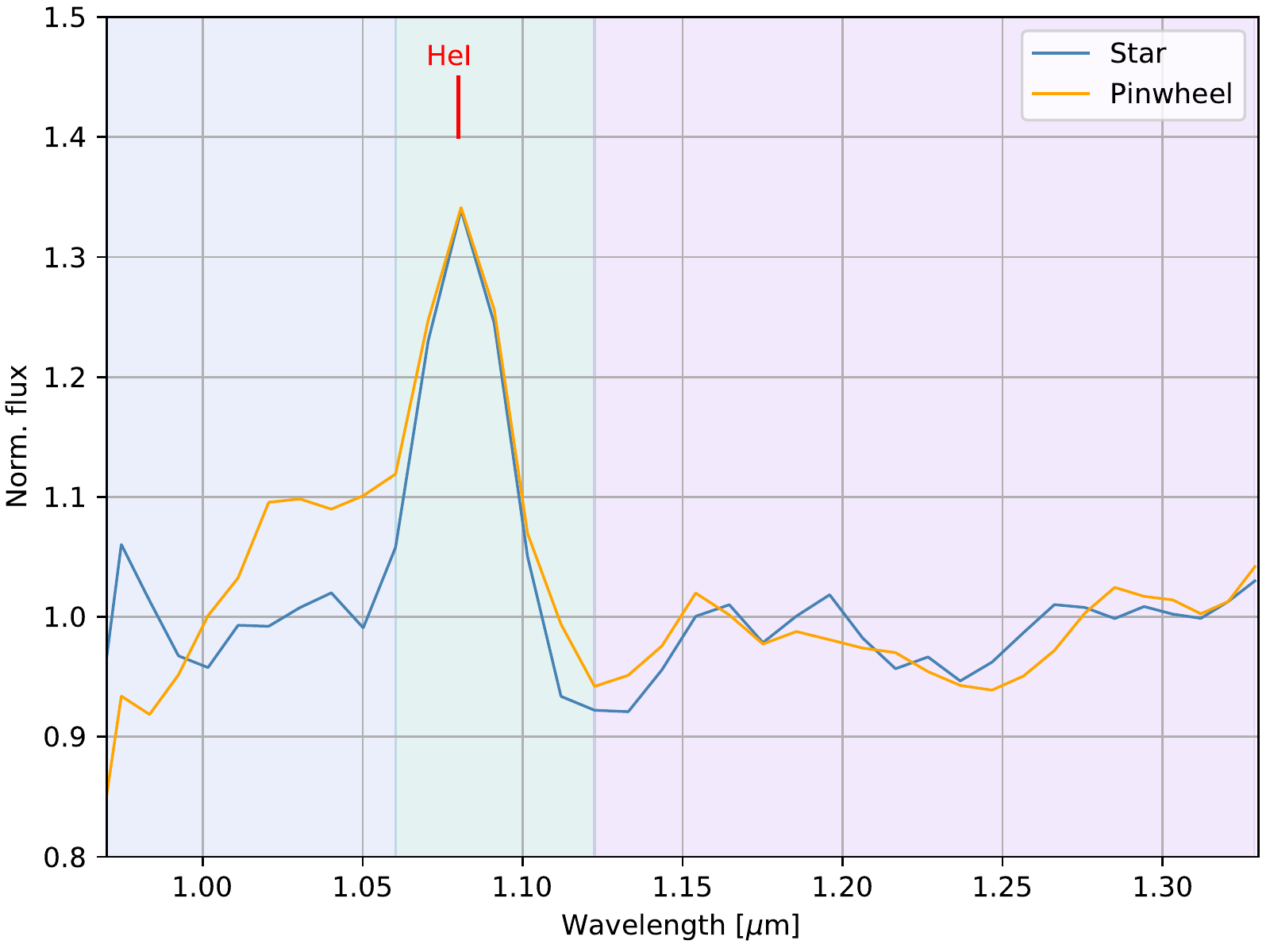} 
		\caption[example] 
	{ \label{fig:spectrum} 
	Spectra of inner (blue) and outer (orange) regions of WR104 in the near infrared (the spectral line at 1.08\,$\mu$m corresponds to HeI). The corresponding images, integrated over the coloured spectral ranges, are shown in Fig.~\ref{fig:IFS_data}.}
\end{figure}

\begin{figure}[htbp]
		\centering
			\includegraphics[width=.48\textwidth]{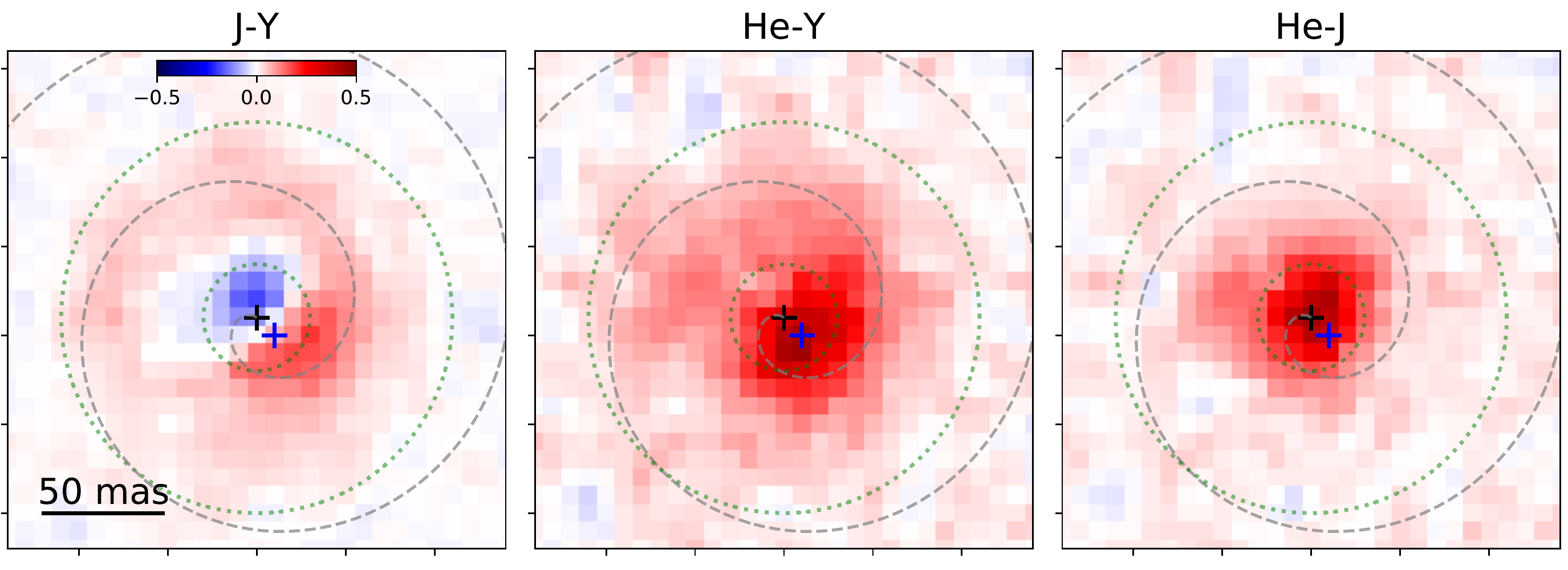} 
		\caption[example] 
		{ \label{fig:color_ifs} Different colour images using Y, HeI, and J bands (blue, green and purple parts of Fig.~\ref{fig:spectrum}). The photocenters of the Y (black cross) and J bands (blue cross) are represented as well as the best fit of the spiral, shown by a grey dashed line. The dashed, green circles both represent regions from where the two spectra have been extracted in Fig.~\ref{fig:spectrum}: from the star (inner region) and the pinwheel (between the two circles).}
\end{figure}

The two spectra of WR104 presented in Fig.~\ref{fig:spectrum} are extracted from the IFS data cube and calibrated using the associated PSF calibrator. We integrate the flux over all 39 spectral channels to give a raw spectrum both for the PSF reference star and WR104. We extracted one spectrum in the inner region (representing the star), and the other in the outer region (representing the pinwheel); these two regions of interest are shown in Fig.~\ref{fig:color_ifs}. 

In the two resulting spectra, we can recognize only one emission line close to 1.08\,$\mu$m, which we identify as a helium line. The same spectral feature appears in both spectra, which is an indication about the reflective nature of the dust emission. Nevertheless, due to the lack of angular resolution of IFS data, we cannot distinguish if this HeI emission came from the Wolf-Rayet star itself or from the hot-shocked plasma into the WCZ (both are contained in the inner region of the image).

The power of IFS data derives from their spatio-spectral nature. We use the spectrum from Fig.~\ref{fig:spectrum} and the IFS data cube to compute three narrow-band images around 1.01\,$\mu$m (Y band image, blue region in Fig.~\ref{fig:spectrum}), 1.09\,$\mu$m (HeI band image, green region in Fig.~\ref{fig:spectrum}), and 1.20\,$\mu$m (J band image, purple region in Fig.~\ref{fig:spectrum}). We normalize all the images by the number of corresponding spectral channels in order to compare their  fluxes. Fig.~\ref{fig:IFS_data} shows the resulting images and the associated reference star for the PSF. Furthermore, we subtract the three resulting images to compute the colour index images (J-Y, He-Y and He-J) presented in Fig.~\ref{fig:color_ifs}. We attribute the main part of the continuum emission to the circumstellar dust and the HeI emission to the central binary star. Therefore, the different colours highlight the different components of the system:
\begin{itemize}
\item In the  (J-Y) image, the dust (red) is separated from the star (blue) and highlights the spiral structure of the system.
\item In the (He-Y) image, most of the observed flux comes from the dust and reveals the dusty environment of the star.
\item In the  (He-J) image, the main part of the observed flux comes from the star and reveals the position of the star.
\end{itemize}
These colour images easily reveal the complex dusty environment of WR104, especially in the case of the WR star, where the HeI line emission dominates this part of the spectrum (i.e.\ 0.95--1.35$\,\mu$m for the IFS-SPHERE instrument).

We also detect a difference in the position of the photocenter in the images for the Y and J bands, offset by $\approx10$\,mas, with a position angle of $\theta_{\rm{offset}}=-140^{\circ}$. This angle does not seem to be correlated with the parallactic angle (-104$^{\circ}$). This offset might be due to a problem of atmospheric dispersion correction. However the SPHERE ADC\footnote{atmospheric dispersion corrector} was designed to allow $<2\,mas$ displacement with wavelength, and the PSF data exhibit an offset of only $\approx1.8$\,mas. Therefore, we conclude that this offset is likely to be real.

\subsection{The companion B of WR104}
\label{compB_study}

Our J band images of WR104 exhibit the presence of a companion at a distance of approximately 1'' from the central binary star. This companion star, hereafter referred to as
``companion B'', was first detected with the HST in 1998 and reported in \citetads{2002ASPC..260..407W}. According to \citetads{2002ASPC..260..407W}, ``the additional component [...] is likely to be physically related to the WR star [...].'' We now have means of verifying this thanks to the new extremely precise, proper motions of stars provided by the Gaia mission \citepads{2017yCat.1340....0Z}.

Because \citetads[]{2002ASPC..260..407W} focused on companion detection rather than astrometry, we retrieve their HST data (3 filters: 336, 439, 555nm) and recompute the astrometry. Our IRDIS-SPHERE data and these HST data were taken 18 years apart. We fit 2D gaussian functions with a standard Levenberg–Marquardt algorithm to represent the central WR+O binary star and the companion B. The resulting vector of separation values d at all epochs and filters is provided in Table~\ref{tab:compa_hst}. The uncertainties given here are the statistical errors scaled to the $\chi^2$ values (using the covariance matrix). We find a small linear motion of the companion B toward WR104 of $4\pm1$ mas between the two epochs and an angular motion compatible with zero.

\begin{table}[htbp]
\centering
        \caption{Astrometry results for the HST and SPHERE data. $\theta$ is the position angle of the companion compared to North (counterclockwise).\label{tab:compa_hst}}
	\renewcommand{\arraystretch}{1.3}
		\begin{tabular}{c c c c}
		\hline
		\hline
		Instrument  & Filters & d [mas] & $\theta$ \\
        HST &F336W & 979.0$\pm$1.5 & 74.6$\pm$0.1\\
        &F439W & 976.8$\pm$1.1 & 74.7$\pm$0.1\\
        &F555W & 978.5$\pm$1.1 & 74.8$\pm$0.1\\
        \hline
        IRDIS & HeI & 975.5$\pm$0.9 & 74.8$\pm$0.1\\
         & CntJ& 973.3$\pm$1.0 & 74.8$\pm$0.1\\
         & CntH& 973.6$\pm$1.3 & 74.4$\pm$0.1\\
         & H2  & 974$\pm$14 & 74.4$\pm$0.8\\
         & FeII& 966.8$\pm$1.3 & 74.5$\pm$0.1\\
         & H3  & 976$\pm$12 & 74.4$\pm$0.7\\
         & H4  & 975$\pm$15 & 74.4$\pm$0.9\\
         & CntK2& 974.8$\pm$2.9 & 75.5$\pm$0.2\\
         & CO & 970.0$\pm$3.3 & 75.2$\pm$0.1\\
        \end{tabular}
\end{table}

The first Gaia data release DR1 \citepads{2016A&A...595A...2G} provides a new and precise position measurement of WR104. The The US Naval Observatory CCD Astrograph Catalog (UCAC5) uses a combination of Gaia DR1 measurements and the Naval Observatory Merged Astrometric Dataset (NOMAD) to compute precise proper motions of millions of stars, including WR104. We retrieved the following UCAC5 proper motions for WR104: pm$_{\rm RA}$=-3.1$\pm$1.0 mas/yr; pm$_{\rm dec}$=-2.4$\pm$1.0 mas/yr \citepads{2017yCat.1340....0Z}. If companion B were a background star (i.e.\ with a negligible proper motion), WR104 should have moved relative to it by $68\pm17$\,mas between the HST and SPHERE observations: such a large motion would have been detectable. This increase is not detected, so companion B is unlikely to be a background star.

\begin{figure}[h]
	\centering
		\includegraphics[width=.48\textwidth]{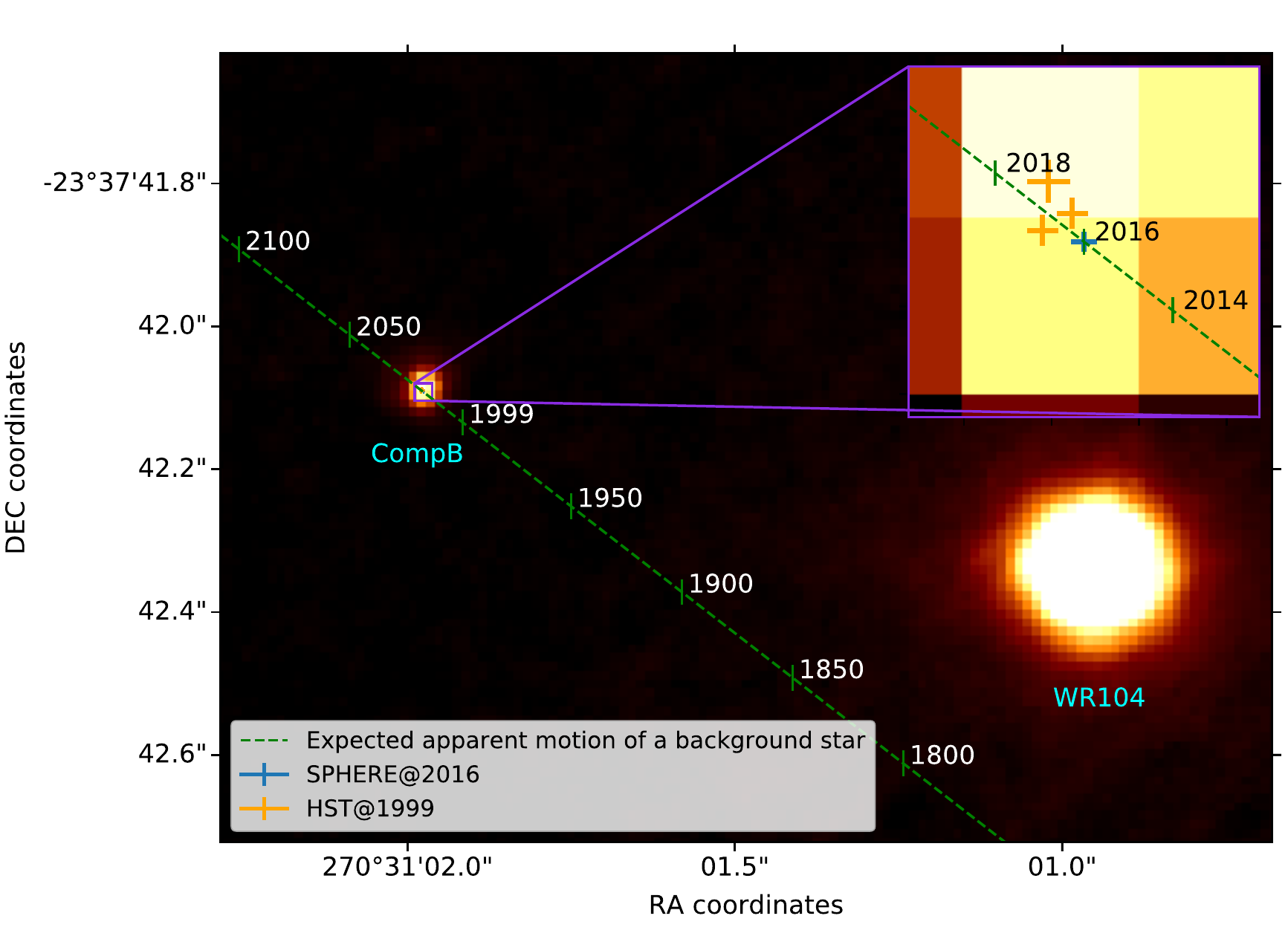}
    \caption{\label{fig:apparent_motionWR104} 
    Measured positions of companion B (CompB) with respect to the position of WR104, set to the epoch of the SPHERE observation. The predicted motion of an hypothetical background star relative to WR104 is presented with a green dashed line. The measured positions of companion B are shown in blue (SPHERE@2016) and orange (HST@1999) crosses in the inset. The widths of the crosses  indicate the uncertainties of the measured positions.}
\end{figure}

In Figure~\ref{fig:apparent_motionWR104}, we show the expected motion of companion B relative to WR104 under the assumption that it is a background star (green dashed line with the position at different dates). In the figure inset, we show the measured position of companion B relative to WR104 made with SPHERE in 2016 (blue cross) and the HST in 1999 (orange crosses). This demonstrates a common proper motion (rate and angle) between WR104 and companion B.

That common proper motion provides an argument in favour of a gravitational link between companion B and the central binary WR+O star, as explained in \citetads{2017ApJS..230...15M}. Therefore, assuming that both stars are at the same distance, we identify companion B and WR104 as a common proper motion binary.

We also consider the effective temperature of the companion B star (defined as $\rm T_B$). To do so, we start by measuring the flux ratio between the WR+O central star and the companion star in all the available filters (see Fig.~\ref{fig:rflux}). We adjust the flux ratio using different atmosphere models to represent the Spectral Energy Density (SED) of the stars with PoWR\footnote{The Potsdam Wolf--Rayet Models, see \url{http://www.astro.physik.uni-potsdam.de/~wrh/PoWR/powrgrid1.php} for details.} models for the WR star \citepads{2015A&A...577A..13S} and Kurucz models for the two companion stars (OB-type inner binary and companion B; \citepads{2004astro.ph..5087C}). The WR+O binary is surrounded by dust (the pinwheel nebula), which is represented by a blackbody with temperature T$_{dust}$. We assume a WC9-subtype for the WR star to obtain the corresponding SED with the PoWR models \citepads{2012A&A...540A.144S}. To select the appropriate Kurucz atmosphere model, we assume the spectral type of the inner OB-type star to be B0.5V as reported by \citetads{2015MNRAS.447.2322R}). We scale the SED of WR+O to match the total luminosity of the binary ($L_{WR+O}=120000\,L_\odot$, see Table~ \ref{tab:stellar_param}). Finally, we compute the flux density using our estimation of the distance ($\rm D=2.58\pm0.12$\,kpc, see Section~\ref{distance}).

We scale the flux density of the pinwheel with the solid angle seen from Earth $\Omega_{dust}$. Then, we compute the flux density of companion B using another scale parameter $\rm r_B$, representing the radius of the star in units of solar radii $\rm r_\odot$.

\begin{figure}[htbp]
	\centering
		\includegraphics[width=.48\textwidth]{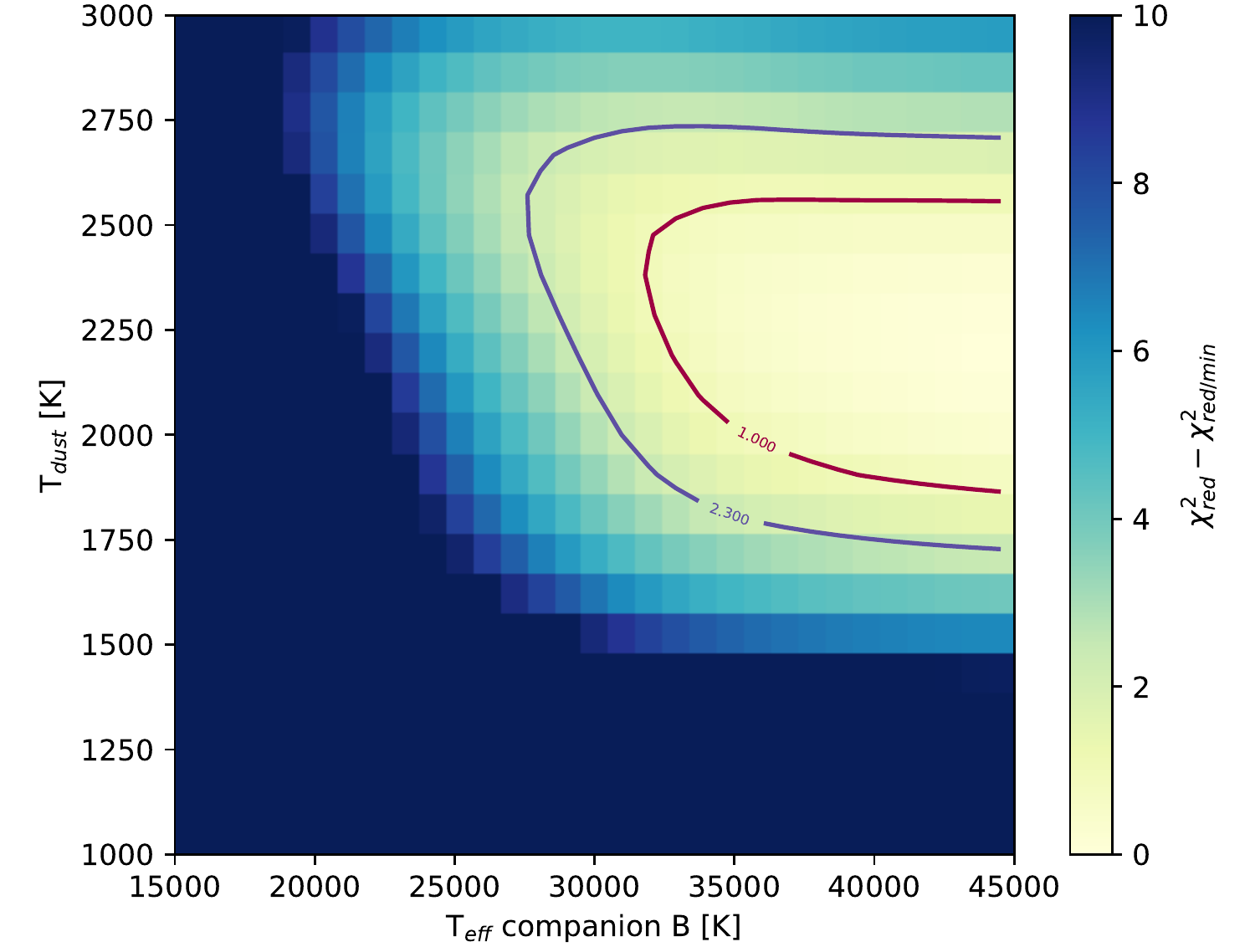} 
    \caption{\label{fig:chi2_map} 
    Reduced $\chi^2$ maps for the flux ratio fitting. Here, we present $\Delta\chi^2_r=\chi^2_r-\chi^2_{r,min}$ and show the respective 1$\sigma$ and 2$\sigma$ confidence intervals with red and purple lines.}
	\vspace{0.5cm}
\end{figure}

We use a Levenberg--Marquardt (LM) minimization of the $\chi^2$ to find the best solution of the flux ratio between companion B and the WR+O+dust component (namely WR104). The fit provides the uncertainties on the scale parameter $\Omega_{dust}$ and the star B radius $\rm r_B$ using the covariance matrix. We used a constant, relative uncertainty on the flux ratio of 3$\%$ (corresponding to an uncertainty of the flux of approximately $2\%$ found with the SPHERE data). Then, we perform a scan of the parameter space $\{\rm T_{B},\rm T_{dust}\}$ to provide a more robust 1$\sigma$ uncertainty. For this study, we set the log(g) parameter in the Kurucz model of companion B to 4.5.\footnote{We set log(g)  to 4.5 to compute the Kurucz models with an effective temperature between 15,000 and 45,000K. Given the high luminosity of companion B compared to the inner binary (factor 4 in luminosity), the companion is likely to be another massive star which is consistent with a high value of log(g).} A precise determination of this parameter is beyond the scope of the determination of the flux ratio. Fig.~\ref{fig:chi2_map} shows the $\chi^2$ map for a range of effective temperatures of the companion B ($T_B$ between 15,000 and 45,000 K) and for the dust temperature ($T_{dust}$ between 1000 and 3000 K). The figure contains the $\chi^2$ map subtracted by the minimum $\chi^2$ value ($\chi^2_{red}=2.3)$ as well as the corresponding 1$\sigma$ and 2$\sigma$ confidence intervals. 

With the wavelengths used in this study (U, B, and V bands from \citetads{2002ASPC..260..407W} and J, H and K bands (this study), we are able to provide a lower limit on the effective temperature of the companion star such that $T_B \geq 33000\,K$ (see Table~\ref{tab:results_compB}). To further constrain the temperature of this companion, we would require additional UV data so as to account for the peak of the emission of the star.
\begin{table}[htbp]
\centering
        \caption{Best fit parameters for the model of WR104 and companion B.\label{tab:results_compB}}
	\renewcommand{\arraystretch}{1.3}
		\begin{tabular}{c c c}
		\hline
		\hline
		Parameters      & Fit ($\chi^2_{red}=2.3$) & Confidence interval\\
        $T_{B}$         & $\geq 33000$ K  & 1$\sigma$ ($\chi^2$ map)\\
        $T_{dust}$      & $2200\pm400$ K & 1$\sigma$ ($\chi^2$ map)\\
        $r_{B}  $  & $\rm 4.3\pm0.5\;r_\odot$   & Covariance matrix \\
        $\Omega_{dust}$  & $3\pm2\times 10^{-16}$ sr & Covariance matrix \\
        $L_{B}$  & 68000 L$_\odot$ & -\\
        Spectral type  & O8V to O5V & -\\
  
        \end{tabular}
\end{table}
\section{Modelling the dust emission in the spiral }\label{sec:model}

\subsection{Simple Archimedean spiral}
\label{distance}
The simplest approach to describe the circumstellar environment of WR104 is to use an Archimedean spiral (also called arithmetic spiral\footnote{The Archimedean spiral has the property that any ray from the origin intersects successive coils of the spiral with a constant separation distance.}) to follow the dust distribution around the central binary star \citepads{1999Natur.398..487T}. The Archimedean spiral takes as hypothesis that the dust is created and accelerated up to a constant velocity and then follows a ballistic trajectory (linear radial motion).

Based on the determined period of $P=241.5\pm0.5$ days, the best fit of the spiral on our images provides an angular speed of $v_{ang}=0.273\pm0.013$ mas/day, which corresponds to the radial outward motion of the dust. It yields a separation of the individual turns of the spiral (``steps'') of $0.273\, \rm{mas/day}\times241.5\,\rm{mas} = 66\,\rm{mas}$.

Previous studies showed that the central binary system has a circular orbit \citepads{2008ApJ...675..698T}. As a consequence, the dust nucleation locus follows the orbital motion of the system. Therefore, the spiral pattern is directly linked to the orbital period of the system and the speed of the expanding dust. \citetads{2008ApJ...675..698T} demonstrated that the wind velocity at large radii (at least at the dust formation locus) is strongly dominated by the WR wind. The latter overwhelms the O star wind and radiative braking.\footnote{Phenomenon occurring in the interaction within the massive binary system, whereby the WR wind is decelerated by the weaker O wind close to the star \citep{1997ApJ...475..786G}.} As such, we consider that the dust expansion velocity is set by the WR wind speed, so $V_{dust} \simeq V_{\infty, WR}$, where $V_{\infty}$ is the terminal wind velocity, with only minor adjustments.

This means that if we know the physical speed of the dust over an orbital period, we can directly determine the distance of the system. Conversely, a precise measurement of the distance can yield a unique measurement of the WR-dominated wind speed. In Table \ref{tab:stellar_param}, we provide all the physical parameters of WR104 found either in the  literature \citepads{1997MNRAS.290L..59C, 2002ApJ...566..399M} or in tables assuming typical values for the WR and O stars \citepads{2012A&A...540A.144S, 2015PASP..127..428F}. These parameters are the temperature ($T$), the stellar mass ($M$), the terminal wind velocity ($V_{\infty}$), the gas mass-loss ($\dot{M}$) and the total luminosity ($L$).


\begin{table}[htbp]
\centering
        \caption{Stellar parameters.\label{tab:stellar_param}}
	\renewcommand{\arraystretch}{1.3}
		\begin{tabular}{c c c}
		\hline
		\hline
		Parameters      &  & References\\
        $T_{WR}$        & 45000 K     &  \citetads{1997MNRAS.290L..59C}\\
        $M_{WR}$        & 10 $M_\odot$ &  \href{http://adsabs.harvard.edu/abs/2012A\%26A...540A.144S}{Sander et al. (2012)}$^1$ \\
        $v_{\infty/WR}$ & 1220\,km/s & \href{http://adsabs.harvard.edu/abs/1992A\%26A...261..503H}{Howarth \& Schmutz (1992)}\\
        $\dot{M}_{WR}$  & 0.8$\times10^{-5}$ M$_{\odot}$/yr & \citet{2002ApJ...566..399M} \\
        $L_{WR}$          & 40000 $L_{\odot}$ & \citetads{1997MNRAS.290L..59C} \\
        \hline
        $T_{OB}$        & 30000 K     & \citet{2004MNRAS.350..565H}\\
        $M_{OB}$        & 20 $M_\odot$ & \citetads{2015PASP..127..428F}$^2$\\
        $v_{\infty/OB}$ & 2000 km/s   & \citet{2004MNRAS.350..565H}\\
        $\dot{M}_{OB}$        & 0.5$\times10^{-7}$ M$_{\odot}$/yr & \citet{2015PASP..127..428F}$^2$\\
        $L_{OB}$        & 80000 $L_{\odot}$ & \citet{2004MNRAS.350..565H}$^3$ \\
        \hline
        \multicolumn{3}{l}{$^1$ See their Table 6, assuming a WC9 subtype.}\\
        \multicolumn{3}{l}{$^2$ See their Table 3, taking $L_{OB}= 80000L_{\odot}$ and $T_{OB}=30000$\,K.}\\
        \multicolumn{3}{l}{$^3$ They use a 1/2 luminosity ratio between WR and O, based}\\
        \multicolumn{3}{l}{     on an assumed V band ratio of 1/2  from \citetads{1997MNRAS.290L..59C}.}\\
        \end{tabular}
\end{table}
Reported terminal velocities of WC9 stars range from 1220\,km/s \citepads{1992A&A...261..503H, 2007ARA&A..45..177C} up to 1600\,km/s \citepads{2012A&A...540A.144S}. Taking the lower value of $1220$\,km$\cdot$s$^{-1}$, we find a distance of $D=2.58\pm0.12$ kpc, which is in good agreement with previous measurements and make WR104 a potential member of the Sgr OB1 stellar association \citepads{1984A&AS...58..163L}. Taking the most recent value of 1600\,km$\cdot$s$^{-1}$ \citepads{2012A&A...540A.144S}, we obtain a distance of $D=3.38\pm0.15$ kpc, that would place WR104 behind the Sgr OB1 association.

Future measurements of distances with Gaia, expected with the data releases DR2 (\citetads{2018arXiv180409365G}, see Sect. \ref{sec:GaiaDR2}) and DR3, will provide a very good estimate of the still unconstrained terminal wind velocity of the WR star, in the case when these WR stars are surrounded by a resolved pinwheel nebula.


\subsection{Phenomenological model}
\label{simplemodel}

To probe more of the system's physical parameters, we update the model from \citetads{2009A&A...507..317M}, available in the public distribution of \texttt{fitOmatic}.\footnote{\url{http://fitomatic.oca.eu}} This updated version relies on physical parameters of the object such as the temperature laws, the dust-formation regions, and the opening angle of the shock. As mentioned in the Introduction, it is the shock caused by the collision between the winds from the WR and O stars that results in the dusty spiral.
This model provides a reasonable match to the flux of the inner part of the spiral.

The model is composed of the pinwheel and an inner binary system with a separation $d_{\rm bin}$ fixed to 1\,mas (2.6\,au at 2.6\,kpc). The pinwheel is constructed with a succession of rings growing linearly with the distance from the central binary and positioned following an Archimedean spiral pattern with a given step. This spiral model is motivated by the hypothesis that the central binary is on a circular orbit and that the dust nucleation occurs at the interface of the shocked winds of the WCZ (\citeads{2012A&A...546A..60L}; \citeads{2016MNRAS.460.3975H}).

The flux contributions of the two components are calculated at different wavelengths using blackbodies. We chose 45,000\,K for the WR star, 30,000\,K for the O star (Table~\ref{tab:stellar_param}) and assumed that each ring of the pinwheel emits as a blackbody at a temperature defined as a decreasing function of the ring's distance to the central system. The beginning of the pinwheel is defined by the parameter $r_{\rm{nuc}}$  corresponding to the dust nucleation location. The temperature T follows a power-law with the nucleation temperature ($T_{\rm{nuc}}$ at the $r_{\rm{nuc}}$ distance) and a power-law index  $q$:
\begin{equation}
	\label{eq:Tlaw}
	T(r) = T_{\rm nuc}\times\left(\frac{r}{r_{\rm nuc}}\right)^{-q}.
\end{equation}

Note that we consider here that the nucleation temperature $T_{\rm{nuc}}$ is identical to the dust-sublimation temperature $T_{\rm{sub}}$. The binary star contribution to the total flux is calculated using the scale parameter ratio$_{\rm star}$, arbitrarily set at a wavelength of $1\,\mu$m.\footnote{ratio$_{\rm star}=1$ means that the pinwheel and the star contribute equally to the SED at 1 micron.} We set the semi-opening angle of the spiral to $\alpha = 17.5^\circ$ as determined by the momentum flux ratio $\eta$ of the winds and the empirical relation from \citetads{1993ApJ...402..271E}:

\begin{equation}
\eta = \frac{\dot{M}_{OB}v_{\infty/OB}}{\dot{M}_{WR}v_{\infty/WR}}, \label{eq:eta}
\end{equation}

\begin{equation}
\alpha = 2.1\left(1-\frac{\eta^{4/5}}{4}\right)\eta^{1/3}, \\ \rm{for}\; 10^{-4} \leq \eta \leq 1.
\end{equation}

\begin{figure}[htbp]
	\centering
		\includegraphics[width=0.48\textwidth]{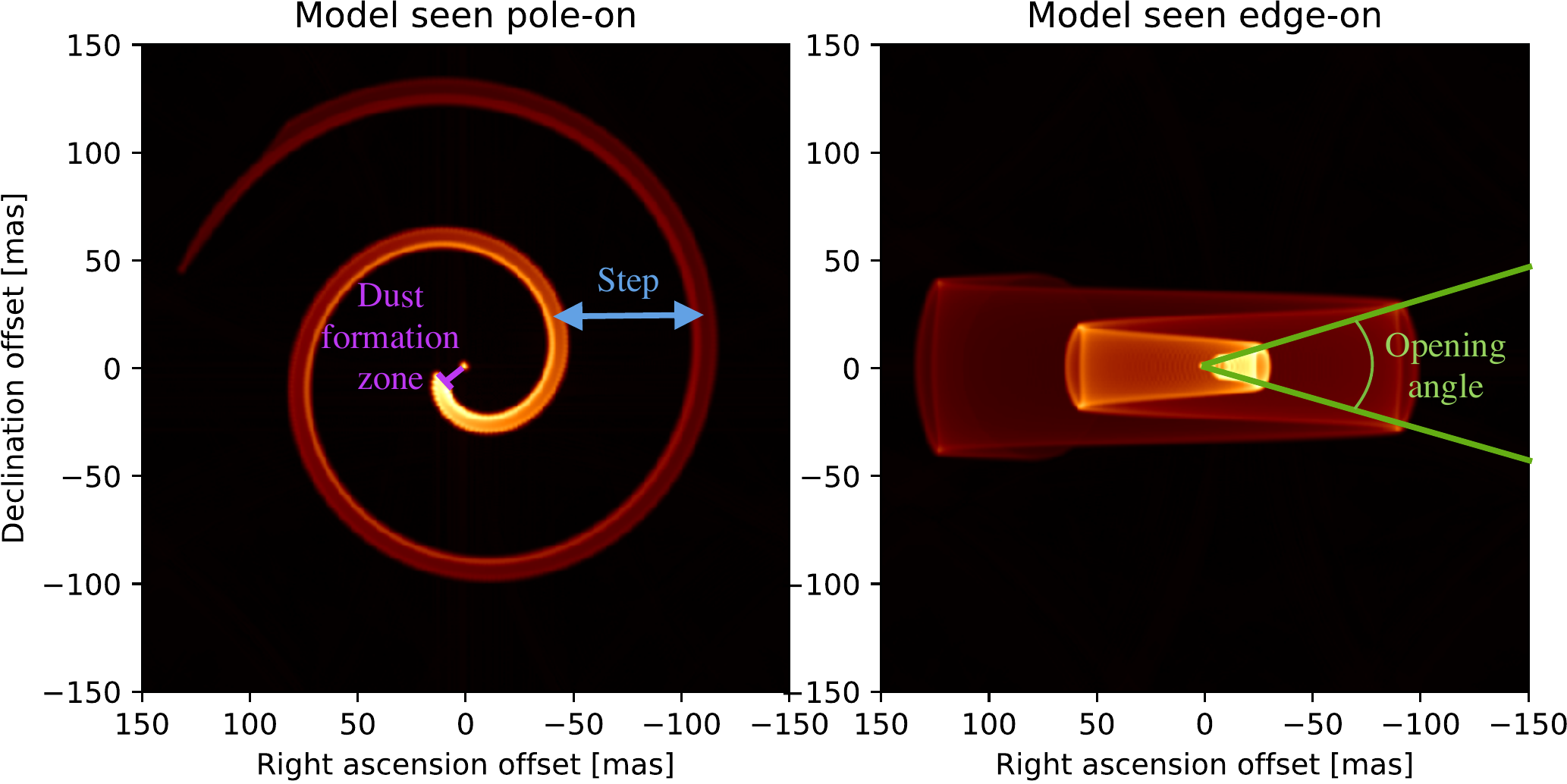} 
	\caption[] 
	{\label{fig:bestmodel} Phenomenological model used in this work. The important parameters are the dust formation zone (purple), opening-angle (green), and step of the spiral (blue). Projection angles on the sky are the same as in \citetads{2009A&A...506L..49M}. The left panel corresponds to the spiral as seen from the Earth (face-on, $i=0^\circ$), and the right panel shows the model seen in its equatorial plane (edge-on, $i=90^\circ$)}	
\end{figure}

We fixed the number of spiral turns to 3.5 to be able to correctly fit the steps and the sky orientation defined by $\theta_0$ (the spiral orientation at the reference date of April 1998). We also set the inclination of the orbital plan to zero ($i=0^\circ$). An illustration of this phenomenological model and its main parameters is presented in Fig.~\ref{fig:bestmodel}.

We use a two-stage model-fitting approach: first we obtain rough values of the parameters  using the deconvolved K-band image only, and then we refine the parameters with the non-deconvolved images. The fitting is a least squares calculation based on the pixel values: 
\begin{equation}
	\chi^2 = \sum_{i} \frac{(O_i-M_i)^2}{e_i^2},
\end{equation}

\noindent where $O_i$ is the i-th pixel's intensity, $e_i$ its associated error, and $M_i$ the prediction from the model. We use a Levenberg--Marquardt (LM) minimization of the $\chi^2$ to find the best solution. To ensure that we have not converged on a local minimum and that our fit is robust, we also scan the parameter space of the relevant parameters (step of the spiral, orientation, x0, y0).

Table~\ref{tab:bestfit_param} presents the best-fit parameters of our phenomenological model. Taking into account the resolution of the VLT in the K band ($\lambda/D=60$\,mas) and the pixel size of IRDIS (12.25 mas), we are able to set a sub-resolution constraint on the step and orientation of the spiral as well as the dust-formation location ($d_{\rm shock}$). By assuming two different sublimation temperatures ($\rm T_{sub}=1500\,K$ as used by \citetads{2004MNRAS.350..565H}; $\rm T_{sub}=2000\,K$ for carbon from \citetads{2011EP&S...63.1067K}), we also derive a new measurement of temperature along the spiral. 

We find better agreement for the higher sublimation temperature of 2000\,K (q$_{2000}=0.45$, $\chi^2 =1.24$). This is consistent with the hypothesis that the dust is composed of amorphous carbon. Indeed, since WR104 is a WC star,  it produces  large amounts of carbon. We  assume that the chemistry of dust nucleation is based  on carbon and not silicates. By fitting the SED of different WR stars of the IRAS catalog (such as WR104), \citetads{1998MNRAS.295..109Z} showed that amorphous carbon seems to be more representative of the dust in the dusty WR star.

The uncertainties are given with a $1\sigma$ confidence level, and each parameter is constrained one by one around a given range of values (e.g.\ $0.3<q<0.5$).  To compare the orientation obtained with SPHERE with those from \citetads{2008ApJ...675..698T}, the phase $\theta_0$ is given at the reference date of April 1998 (MJD = 57512). Our two epochs allow us to confirm the 241.5\,day orbital period of the pinwheel.

\begin{table}[htbp]
\centering
        \caption{\label{tab:bestfit_param} Best fit parameters of the phenomenological model. The first reduced $\chi^2$ corresponds to the the dust sublimation temperature of 1500\,K, the second to 2000\,K.}
	\renewcommand{\arraystretch}{1.3}
		\begin{tabular}{c c}
		\hline
		\hline
		Parameter      & Fit ($\chi^2 = 1.34/1.24^*)$\\
        step        &  66 $\pm$ 3 mas\\
        r$_{nuc}$ &  12 $\pm$ 2 mas\\
        q$_{1500}$ &  0.37 $\pm$ 0.03\\
        q$_{2000}^*$ &  0.45 $\pm$ 0.03\\
        $\theta_0$ & 260 $\pm$ 2 deg\\
        \hline
        $\alpha$ & 17.5 deg (fixed)\\
        turns & 3.5 (fixed)\\
        ratio$_{star}$ & 1 (fixed)\\
      	$d_{\rm bin}$ & 1 mas (fixed)\\
        i & 0 deg (fixed)\\
        \end{tabular}
\end{table}

\subsection{Radiative transfer modelling}
\label{sec:radmodels}

Although a complete radiative transfer model of WR104 is beyond the scope of this paper \citepads[see the study by][]{2004MNRAS.350..565H}, a qualitative comparison based on radiative transfer is necessary to further interpret the new WR104 data presented here.

To do so, we compare our images with a 3D axisymetric radiative transfer model based on a geometric model made of consecutive rings. These rings represent the consecutive coils of the dusty Archimedean spiral at a given azimuth. The corresponding 3D density grid is then fed into the well-established publicly available code RADMC3D \citepads{2012ascl.soft02015D}, which computes the self-consistency of temperature distributions and produces mock images.

RADMC3D uses a Monte-Carlo method and launches photon packets from the central star into the dust density grid in order to compute the dust temperature at each point of the grid. The scattering source function is then computed at each wavelength through an additional Monte Carlo run. We assumed an isotropic scattering. A last step uses ray tracing to compute the mock images.

The near infrared (NIR) continuum emission we are modelling is dominated by the dust. Since we do not aim for a detailed fit of the spectral features already done elsewhere, our approach is to assume realistic standard opacity laws. For the spiral, we expect a sub-micron grain size due to the rapid growth of the dust nuclei \citepads{1998MNRAS.295..109Z}, so we assume a pure amorphous carbon composition with a grain size of 0.1\,$\mu$m. In this model, the total dust mass is degenerate with the grain size, and results for the total mass are valid only for the  grain size considered.

\begin{figure}[htbp]
	\centering
		\includegraphics[width=.46\textwidth]{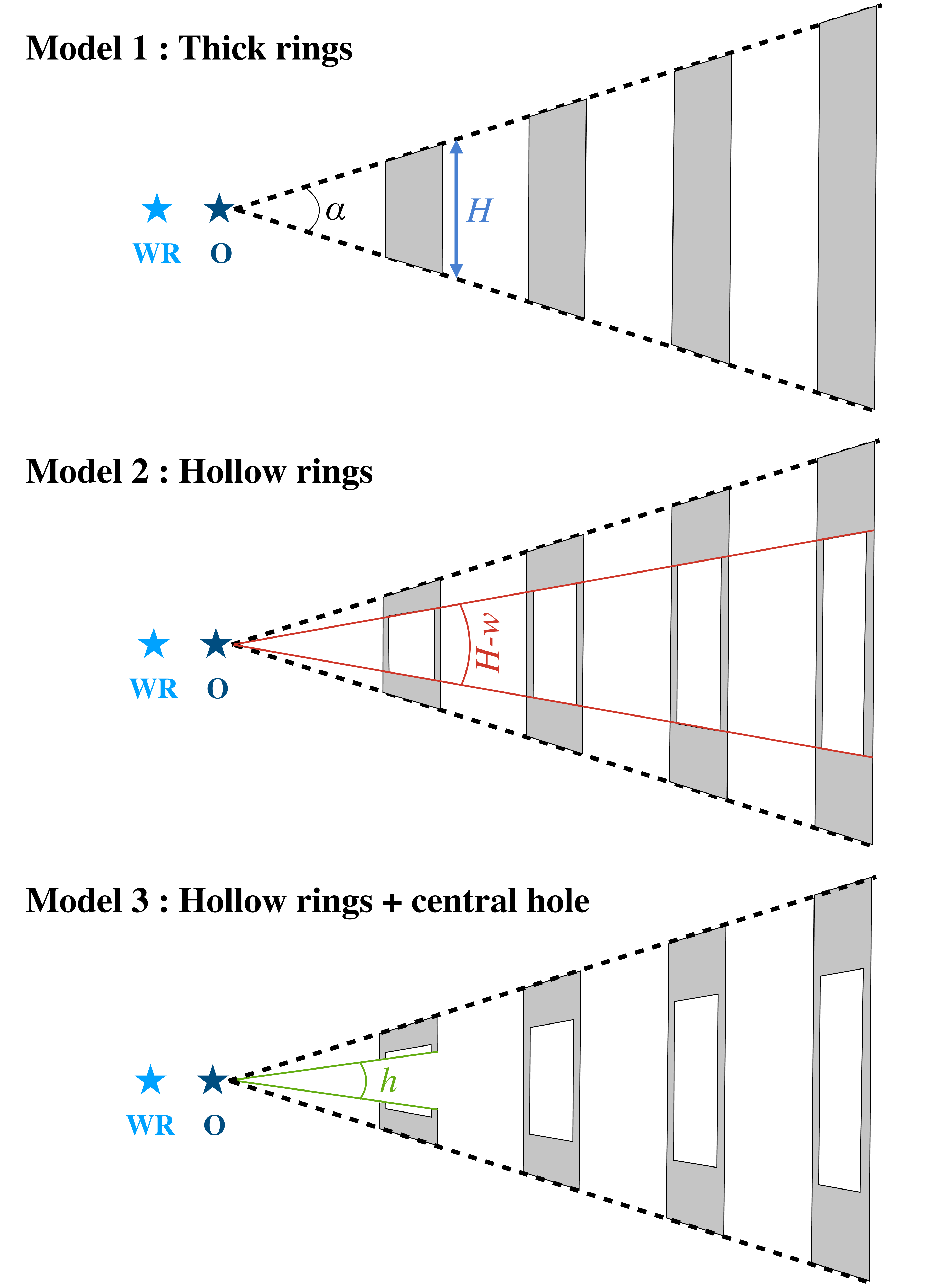} 
	\caption[RADMC3D 2D radiative transfer models used in this paper.] 
	{\label{fig:radmc_model}The different dust density distributions considered in our radiative transfer models. {\bf From top to bottom (edge-on view of the rings model delineated by dashed lines):} full rings, hollow rings, and hollow rings with a central hole. See text for details.}	
\end{figure}

Three dust density models are presented in Fig.~\ref{fig:radmc_model}, with each one corresponding to a physical hypothesis about the dust nucleation:
\begin{itemize}
\item Model 1 consists of uniformly thick rings of linearly-increasing height $H$ separated by the spiral step and an opening angle $\alpha$ set by the WCZ opening angle. The theoretical basis is that the dust is formed downstream from the reconfinement shock behind the O star, where the dust density can reach very high values for low temperatures \citepads{2012A&A...546A..60L}. The dust then follows a ballistic trajectory, and its size grows linearly with the distance, which yields a uniform density in the rings.
\item Model 2 consists of hollow rings, with the width of the walls set by $w$ (expressed as a fraction of the total ring size $H$). Such a model is associated with the hypothesis of dust nucleation at the interface of the shock. The wind collision interface far from the central binary star offers ideal conditions to reach critical densities. This makes dust nucleation possible especially where the mixing between the carbon-rich wind of the WR component and the H-rich wind of the O component becomes significant \citepads{2016MNRAS.460.3975H}. As in Model 1, the dust rings will then grow linearly to create successive coils with a constant fraction of over-density in the walls of the rings.
\item Model 3 also consists of hollow rings. This time the first ring also  has a hole of size $h$ (expressed as a fraction of the ring height) in both walls facing the central stellar source. This setup simulates variations or breaks in the dust density, which could stem from the turbulence inside the shocked region. In this case, dust formation is not completely uniform. \citet{2012A&A...546A..60L} and \citet{2016MNRAS.460.3975H} discussed the possibility of hydrodynamical instabilities at work in the WCZ, which would then create a non-uniform density distribution. For instance, thin shell instabilities occur when cooling is important so that the shocked zone narrows to a thin layer, which is easily perturbed \citepads{1994ApJ...428..186V}. Such instabilities provoke strong distortions of the whole colliding region (\citeads{2009MNRAS.396.1743P}; \citeads{2011A&A...527A...3V}). Therefore, if the dust is formed in this turbulent region, the resulting density distribution will be non-uniform with over- and under-densities.
\end{itemize}

For the three types of models, the masses M of the rings are independent of distance and calculated using the period of the spiral P, the mass-loss rate $\dot{M}$ of the dominant\footnote{According to the previously reported values of the mass-loss rates (Table~\ref{tab:stellar_param}),  the mass-loss of the OB component is negligible compared to the WR mass-loss. The resulting total dust mass differs only by $0.6\%$ when neglecting the OB component.} WC9 component, and a dust-to-gas ratio $\xi$: 

\begin{equation}
M = \xi \dot{M} \frac{P}{365.25}.
\end{equation}

We compare our model intensity profiles convolved with the PSF to the observed radial intensity profiles in the J, H, and K bands at a given azimuth, which is set such that the first ring is positioned at 30 mas from the central binary star. To avoid any perturbation caused by the bad quality of the PSF reference star (low dynamic range), we make the comparison between 30 and 300 mas.

We test a range of dust-to-gas ratios (from $\xi=0.1\%$ to $\xi=50\%$) on the three models, with a fixed value of $w=10\%$ and $h=30\%$. For each model, we compute a $\chi^2$ criterion\footnote{We assume that the uncertainties in the data are dominated by the speckle noise. The errors are determined by computing the unbiased standard deviation  for each pixel (Eq.~\ref{Eq:unbiased_STD}).} for comparison.

For all models considered, we cannot satisfactorily reproduce the radial profile obtained with SPHERE. Particularly in the J band, the emission from the dusty rings, including thermal emission and scattering, is negligible relative to the star. Yet, the WR104 images in the J band are definitely resolved with a significant fraction of dust emission (Fig.~\ref{fig:all_red}, 2nd row at 1.21\,$\mu$m). For this reason, we do not include the J band in our comparison.  

\begin{figure}[h]
	\centering
		\includegraphics[width=.44\textwidth]{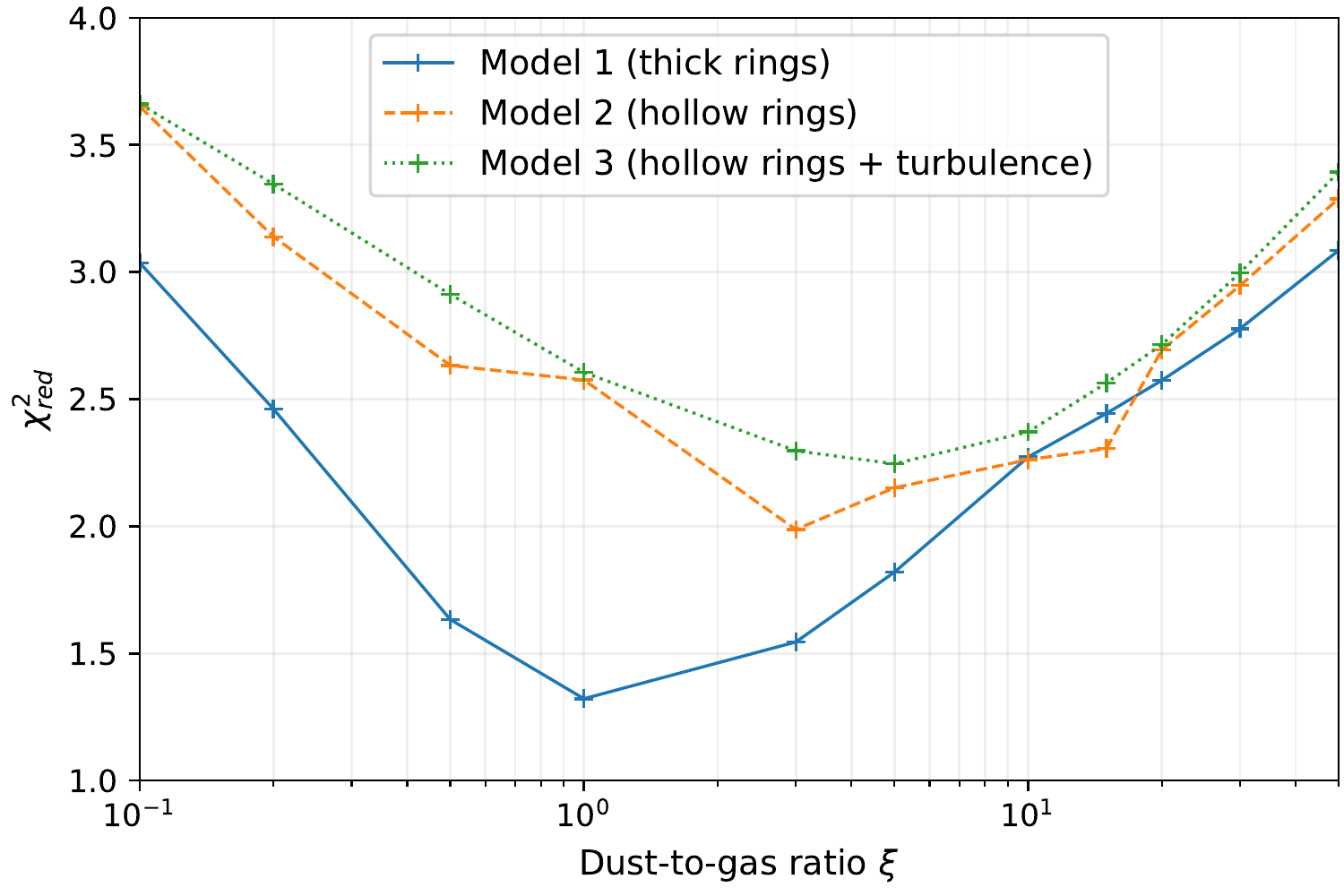} 
	\caption[] 
	{\label{fig:radmc_chi2_allmodels}Comparison of the $\chi^2_{\rm red}$ for the three models considered: Model 1 (solid blue line), Model 2 (dashed orange line), and Model 3 (dotted green line).}	
\end{figure}

\begin{figure}[h]
	\centering
		\includegraphics[width=.48\textwidth]{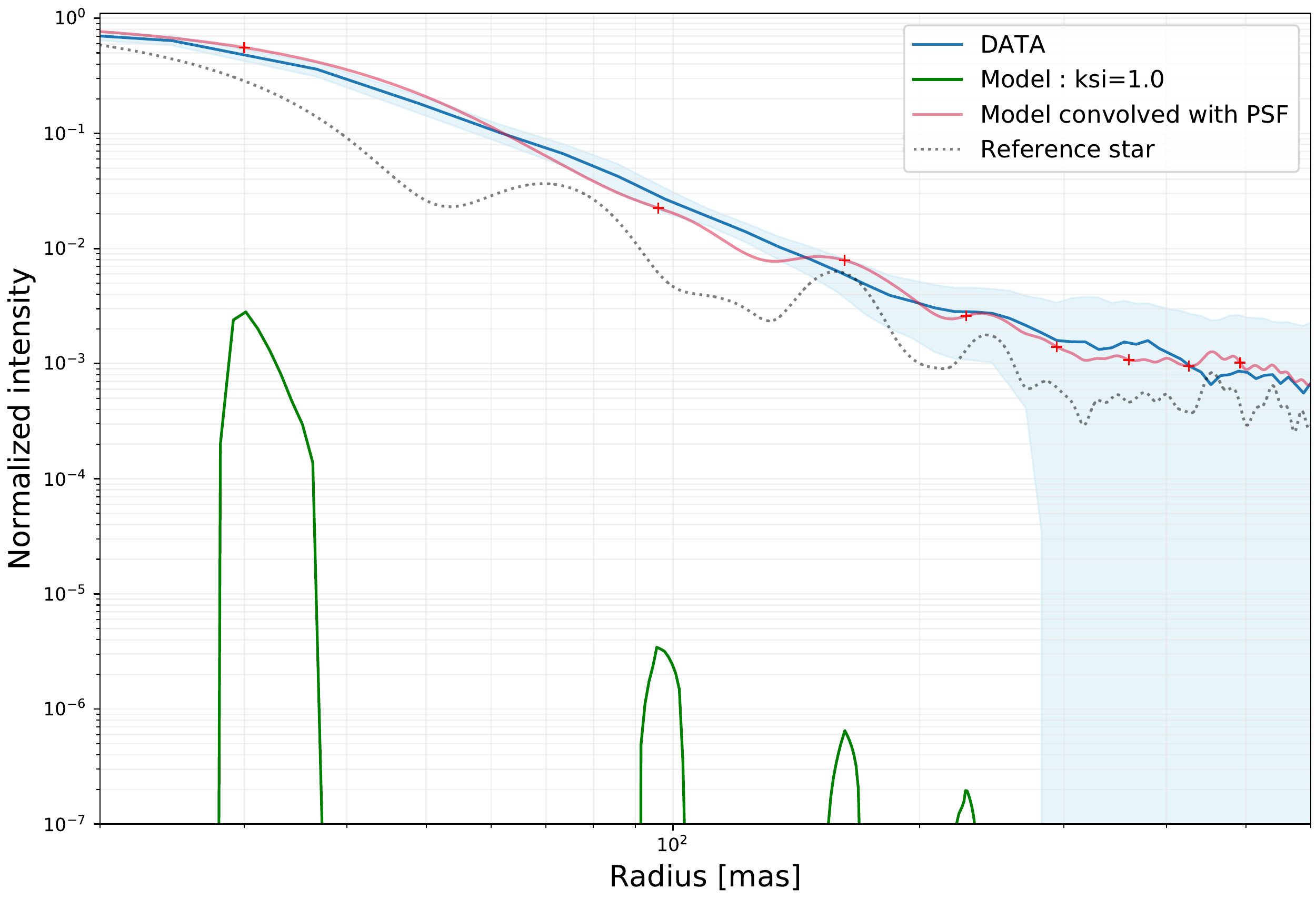} 
	\caption[] 
	{\label{fig:bestmodel_radmc} The comparison between the best model (Model 1, $\xi=1\%$) and the H band data. We represent the radial profile for the: model (green),  reference star (dashed grey),  model convolved with the PSF reference star (red), and the data (blue). The comparison points are represented with red crosses.}
\end{figure}

Considering the H and K band data only, the resulting $\chi^2_{\rm red}$ computation for all considered dust-to-gas ratios is shown in Fig.\ref{fig:radmc_chi2_allmodels}. It appears that Model 1 (thick rings) provides the best match ($\chi^2_{\rm red}=1.31$), with $\xi=1\%$. Fig.~\ref{fig:bestmodel_radmc} shows the corresponding observed and modelled radial profiles in the H band. 

As shown in Fig.~\ref{fig:radmc_chi2_allmodels}, the agreement of Models 2 and 3 with the H and K band data is not as good. Nevertheless, the best match is obtained for $\xi=5\%$ with $\chi^2_{\rm red}=2.15$ for Model 2 and $\chi^2_{\rm red}=2.25$ for Model 3. 

\section{Discussion}\label{sec:discussion}

\subsection{Curvilinear profile and shadowing effect}
\label{sec:shadow}

In section \ref{simplemodel}, we presented the different results obtained from the fit of our phenomenological model. We showed that our model is in good agreement with the K band non-deconvolved data ($\chi^2_{\rm red}=1.24)$. Our model was based on  a power-law decrease of the temperature with a constant index ($q_{\rm 2000} = 0.45\pm0.03$) and a dust sublimation temperature of 2000\,K. This value is very close to the theoretical power-law index of 0.5 corresponding to an optically thin dust distribution \citepads{2010ARA&A..48..205D}. The continuous decrease in temperature implies a continuous curvilinear intensity profile along the spiral. This is in contradiction with the rapid drop in temperature, and correspondingly in intensity, detected after the first spiral turn by \citet{2004MNRAS.350..565H}. They showed that an optically thick first turn could explain such a temperature drop by shadowing the rest of the spiral. With complementary data, \citet{2008ApJ...675..698T} derived a curvilinear intensity profile, which also exhibited an important drop after the first coil.

On this basis, we further investigate the possibility of shadowing effects in this system. For this purpose, we create an artificial gap factor in the spiral intensity profile of our best phenomenological model. This gap represents the shadow cast by the first turn of the spiral onto the second and mimics a significantly optically thick first turn. Following \citetads{2008ApJ...675..698T}, we set this gap factor to 10. We compare the K band data with the curvilinear profile of the model with and without gap over three revolutions.

\begin{figure}[htbp]
	\centering
		\includegraphics[width=.45\textwidth]{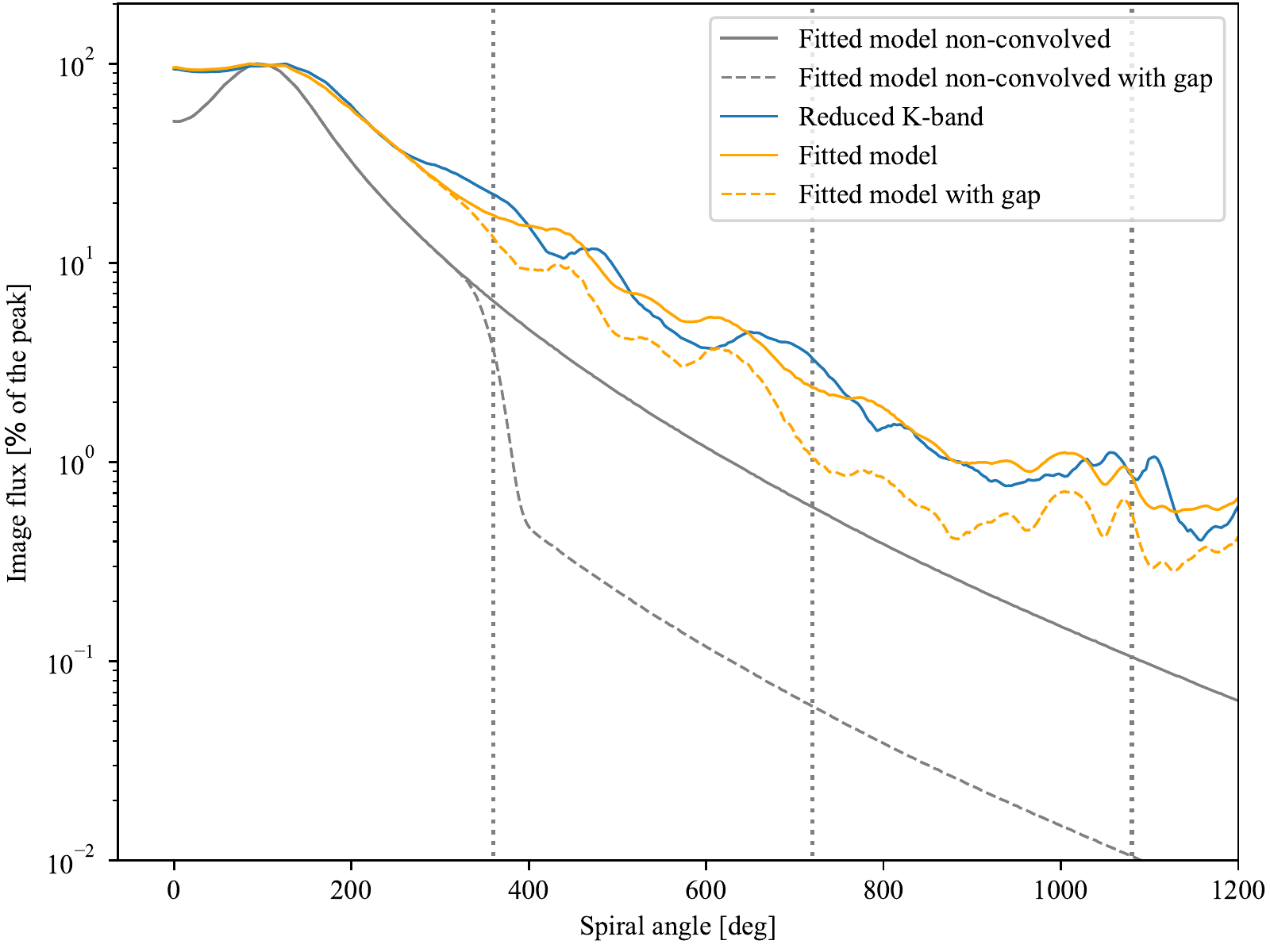} 
	\caption[]{\label{fig:bestfit3turn} Comparison of the curvilinear flux in the K band over 3 orbits. We represent the best model with the gap (dashed line) and without it (solid line) after the first turn. There is  no apparent evidence of the gap in the data ($\chi^2_{gap}=4.3$, $\chi^2_{nogap}=1.5$).}
\end{figure}

In Fig.~\ref{fig:bestfit3turn}, we show both models: with an optically thin first turn (solid grey line) and  with an optically thick first turn (dashed grey line). Then we convolve these models with the PSF reference star (orange lines) and compute the $\chi^2_{\rm red}$ for the models with and without the shadowing effect. The model with a gap seems to be less representative ($\chi^2_{gap}=4.3$ compared to $\chi^2_{nogap}=1.5$).

This conclusion is in apparent contradiction with the results of \citetads{2004MNRAS.350..565H}. They considered an optically thick inner part of the spiral that would produce steps in the curvilinear profile; such steps were detected afterwards by \citetads{2008ApJ...675..698T}. The previous studies were based on image restoration techniques that are known to produce artefacts in the flux's spatial distribution. Here, we based our analysis on direct imaging data with extreme adaptive optics and had a Strehl ratio of 0.87, which also has its flaws. Both image restoration and direct imaging were used at the limit of angular resolution. In our case, the PSF convolution significantly smooths possible intensity variations in the curvilinear profile. Images with a better resolution will be needed to definitely reach a conclusion on this aspect (e.g.\ images from the European Southern Observatory's new MATISSE spectro-interferometer or the planned Extremely Large Telescope (ELT)).

\subsection{Radiative transfer and dust mass}
\label{sec:radmc_discuss}

Our first conclusion from the radiative transfer model is that in the J band, we overestimate the stellar contribution with respect to the observations. This suggests the presence of circumstellar dust along the line of sight, which could obscure the star mostly in the J band, and also very likely in the V band. At this point of the study, we cannot infer a geometry of this dust and can only consider uniform attenuation. This conclusion is supported by the fact that the WR104 system has regular eclipses in the optical (up to 3 mag, \citeads{2014MNRAS.445.1253W}).

Considering the H and K band data only, the comparable value of the different $\chi^2_{\rm red}$  do not allow us to clearly discriminate between the three models of the dust geometry.  Fig.~\ref{fig:bestmodel_radmc} shows the strong impact of the convolution of the PSF on the modelled intensity profile. All differences between the three models are significantly attenuated by the convolution process and cannot be satisfactorily constrained by our data set. This is partly due to the relatively bad quality of our reference star, especially in the K band. It appears that the chosen reference star presents an infrared excess in its Spectral Energy Distribution (SED), which is indicative of circumstellar material. The reference star also presents a very low dynamic range with a poor signal-to-noise ratio, which is probably due to the unoptimized choice of neutral density and exposure time during the observation.

Nevertheless, a trend emerges from the study of the dust-to-gas ratio. All models favour an intermediate dust-to-gas ratio, between 1 and 10$\%$. This translates into a  masses of the dust rings between $5.3\times 10^{-8}M_{\odot}$ and $5.3\times 10^{-7}M_{\odot}$. For comparison, \citeads{2004MNRAS.350..565H} set the mass-loss rate of $\dot{M} = 3\times10^{-5}M_{\odot}$/\,yr in their model and determined a dust-creation rate of $\dot{M} = 8\times10^{-7}$ $M_{\odot}$/\,yr, which yields a dust-to-gas ratio of $\xi_{\rm Harries} = 2.7\%$. This leads to a ring mass of $5.3\times10^{-7}M_{\odot}$. Our results are compatible with these values but seem to suggest a less massive  dusty environment  than inferred by \citetads{2004MNRAS.350..565H}. This raises the question of the efficiency of dust nucleation processes around WR stars.

\subsection{The third star}
\label{sec:companion}
The  third ``companion B'' star located at $\approx$1'' from the central binary is confirmed to be a hot star gravitationally-linked to the inner binary. This confirms that WR104 is at least a hierarchical triple massive system.

Assuming that companion B is orbiting in the same plane as the central binary star, i.e.\ in the plane of the sky (the inclination of WR104 is close to 0), we can estimate the physical separation between companion B and WR104. Taking into account the distance of the system $D=2.58\pm0.12$ kpc (see Sect.~\ref{distance}) and assuming a circular orbit (e=0), the semi-major axis would be $a\approx 2480\pm120$ au.\footnote{This semi-major axis corresponds to the reported distance of 975.5 mas in the HeI filter between companion B and WR104.}

To estimate the orbital period of this triple system, we can make assumptions on the stellar masses of the two O star components (companion B and the OB companion of the WR star). The luminosity and the measured effective temperature \citepads{1997MNRAS.290L..59C} yield an inferred a mass of M$_{\rm OB}=20M_{\odot}$ and a corresponding age of approximately 7 Myr along the main sequence (CMFGEN models, \citeads{2015PASP..127..428F}). According to \citetads{2002ASPC..260..407W} and our study of the effective temperature of  companion B (Sect.~\ref{compB_study}), the companion is very likely another massive OB type star. Therefore, we also set the mass of the companion B at M$_{\rm OB}=20M_{\odot}$. According to the WR star's spectral type as a WC9 \citepads{2012A&A...540A.144S}, we set its mass  to 
M$_{\rm WR}=10M_{\odot}$. We treat the central binary star (M$_{\rm WR+O}=30M_{\odot}$) as a single star and compute the orbital period of the companion B star as 
\begin{equation}
P = \sqrt{\frac{4\pi^2a^3}{G (M_{WR+O}+M_{OB})}} \approx 17000\pm1300\,\rm{days}.
\end{equation}

The orbital plane of the central binary star is close to the plane of the sky ($\leq 16^\circ$). The projected separation between the central binary and companion B is 977\,mas, and the distance between two spiral steps is measured to be 66\,mas. If, as we assumed at the beginning of this section, the tertiary companion B star orbits around WR104 in the same orbital plane as the central binary, we expect it to eventually encounter the fifteenth outer dust coil, produced 10 years ago. Such an encounter might be seen either with hot dust around companion B or a wealth of other phenomena (X-rays, bow shock, etc.)

We do not detect hot dust (infrared excess) around companion B. This means that it may lie outside of the plane of the binary. Based on the estimate of the opening angle of the wind collision zone of $35^\circ$ (see Eq.~\ref{eq:eta}), we find that ``B'' may located beyond $35^\circ$. 

A misaligned set of orbital planes in triple systems has already been observed in other massive stars systems, like Algol \citepads{2012ApJ...752...20B} and $\sigma$ Ori \citepads{2016AJ....152..213S}. Recent N-body numerical simulations report that the non-alignment of triple systems is very common, especially for large projected separations (>1000\,AU, \citeads{2017ApJ...844..103T}). The formation of multiple massive stellar systems is still not fully understood. Multiple isolated systems are probably the result of the collapse of massive cores \citepads{2009Sci...323..754K}. The gravitational instabilities occurring during the formation cause protostellar disk fragmentation, allowing massive multiple systems to form and to overcome the UV radiation pressure barrier (\citeads{1987ApJ...319..850W}, \citeads{2017arXiv170600118M}). Some hypotheses are favoured to explain the misalignment of triple systems: possible accretion of gas with randomly aligned angular momentum at the epoch of star formation, or dynamical processes such as the eccentric Kozai mechanism \citepads{2016MNRAS.456.4219A}. 

Hierarchical triple systems are possible candidates to be at the origin of black holes or neutron star mergers \citep{2017ApJ...836...39S}. Therefore, the WR104 system could be a progenitor of future compact object merger and gravitational wave emission.


\subsection{Gaia DR2 distance and wind velocity}
\label{sec:GaiaDR2}

As discussed in Sect.~\ref{sec:model}, an interesting characteristic of the pinwheel nebula is the possibility to obtain a distance estimation using some hypotheses about the wind velocity. Now, if we are able to find the distance another way (parallaxes for instance), we can retrieve the dust velocity of the pinwheel and the associated wind speed. We used the last distance determination provided by the Gaia DR2 \citepads{2018arXiv180409365G}. With the Gaia parallax ($\rm plx_{G} = 0.2431\pm0.0988\,mas$), we can naively compute the new distance of the WR104 system as $\rm D_{Gaia}=1/plx_{G}=4.11\pm1.67$ kpc, which would place WR104 further than expected, but is still consistent with the previous values. Therefore, if we compute the corresponding dust speed, we find $\rm V_{dust} = 1945 \pm 795\,km/s$ which is compatible but higher than previously reported.

\citetads{2018arXiv180410121B} used a more sophisticated method to compute the geometrical distance using the Gaia DR2 catalog. The non-linearity of the transformation and the asymmetry of the resulting probability distribution does not allow a simple inversion of the Gaia parallaxes. They used a Bayesian procedure and a galactic model to give a more accurate estimate of the distance with a 68$\%$ confidence interval. This method give a distance for WR104 of $\rm D_{Bailer}=3.64 \,{+1.92/-1.02}\,kpc$, which is still higher than expected but nevertheless compatible with our distance estimate ($\rm D = 2.58\pm0.12\,kpc$; Sect.~\ref{distance}). This latest distance measurement allows us to infer the associated dust velocity as $\rm V_{dust} = 1721 \pm 911\,km/s$. We used the upper limit of the distance uncertainty ($\rm 1.92\,kpc$) to propagate the error on the speed.
This velocity is higher than the terminal wind speed of the Wolf-Rayet component, usually used as the dust velocity at the dust nucleation locus \citepads{2008ApJ...675..698T}. The relatively high uncertainty of the Gaia DR2 prevents us from drawing a conclusion about the exact dust velocity. Nevertheless, the trend is towards a higher speed than expected, which could be explained by dust nucleation closer to the star or more acceleration by the wind of the OB star. This raises the question about the still unconstrained dust formation locus as well as the velocity of dust launch.

Nevertheless, some warnings were reported concerning the Gaia DR2 parallax measurements. In particular, in the case where the source is bright (G<13, \citepads{2018arXiv180409366L}), extended, distant ($\rm D>100\,pc$), and lacking a Gaia measurement of radial velocity. Since WR104 fulfils all these criteria, we are cautious with this naive distance estimate.

\section{Conclusions}

With direct imaging using the SPHERE and VISIR instruments on the VLT, we have confirmed the spiral structure of the WR104 system for the first time. We probed the extension of the dust spiral over 15 revolutions with SPHERE and 30 with VISIR  with an unprecedented dynamic range. This corresponds to the history of mass-loss in the last 20 years. Further, we determined a step of the spiral of $66\pm3$ mas, leading to a distance of the system of $2.58\pm0.12$ kpc, thus refining the previous estimate by \citetads{2008ApJ...675..698T}.

Based on the IFS data, and the model fitting of IRDIS data, we confirmed the presence of the dust formation zone approximately 12 mas away from the central binary. The determination of the dust formation region is done at the performance limit of the SPHERE instrument and needs to be confirmed with the higher angular resolution capabilities of interferometric instruments. In the future, the next generation of interferometric instruments, such as MATISSE (\citeads{2014Msngr.157....5L}, \citeads{2016SPIE.9907E..0AM}), will be able to reveal the spiral shape of WR104 at its peak-emission (L-band) and with unprecedented spatial resolution (i.e.\ 4 mas).

We also confirm that the third component of the system, named ``companion B'', discovered with the HST \citepads{2002ASPC..260..407W}, could be gravitationally bound to the inner binary. This companion likely orbits in a different plane compared to the central binary and provides information on the formation of the triple system (competitive or core accretion). 

We created an updated version of the phenomenological model presented in \citetads{2009A&A...506L..49M} and the curvilinear profile of the pinwheel to show that the shadowing effect previously reported does not seem to occur around WR104. This conclusion needs to be confirmed with a new K band high-dynamic range measurement, which would allow the system to be followed during a longer time lapse.

The radiative transfer models used in this work seems to favour a dust-to-gas ratio between 1$\%$ and 10$\%$. Unfortunately, our dataset does not allow us to distinguish between the different hypotheses of the process of dust nucleation. This highlights the need to acquire a new SPHERE dataset with larger bandwidth filters (to improve the signal-to-noise ratio) and a better reference star in the J and K bands (4Sgr appears to be a good candidate considering the better quality of the H band).

Furthermore, to make conclusions about the flow dynamics as well as the dust formation process and its exact location, we need more sophisticated models, including coupling  hydrodynamical models with the radiative transfer processes. This accurate determination is only achievable with a thorough comparison between high-level models and new interferometric data.

\begin{acknowledgements}
This work  uses the SPHERE Data Centre, jointly operated by OSUG/IPAG (Grenoble), PYTHEAS/LAM/CeSAM (Marseille), OCA/Lagrange (Nice), and Observatoire de Paris/LESIA (Paris) and supported by a grant from Labex OSUG@2020 (Investissements d’Avenir – ANR10 LABX56).
We acknowledge support from the CNRS, University of Nice Sophia Antipolis, and the Observatoire de la C\^ote d'Azur. We used NASA’s Astrophysics Data System Bibliographic Services. This research made use of Astropy,\footnote{Available at \url{http://www.astropy.org}} a community-developed core Python package for astronomy \citepads{2013A&A...558A..33A}.  Support for A.L. was provided by an Alfred P. Sloan Research Fellowship, NASA ATP Grant NNX14AH35G, and NSF Collaborative Research Grant 1715847 and CAREER grant 1455342.

\end{acknowledgements}

\bibliographystyle{aa}	
\bibliography{aa_biblio} 
\addcontentsline{toc}{chapter}{Bibliographie}

\appendix
\onecolumn

\section{Filter transmissions}

\begin{figure}[htbp]
	\centering
		\includegraphics[width=.96\textwidth]{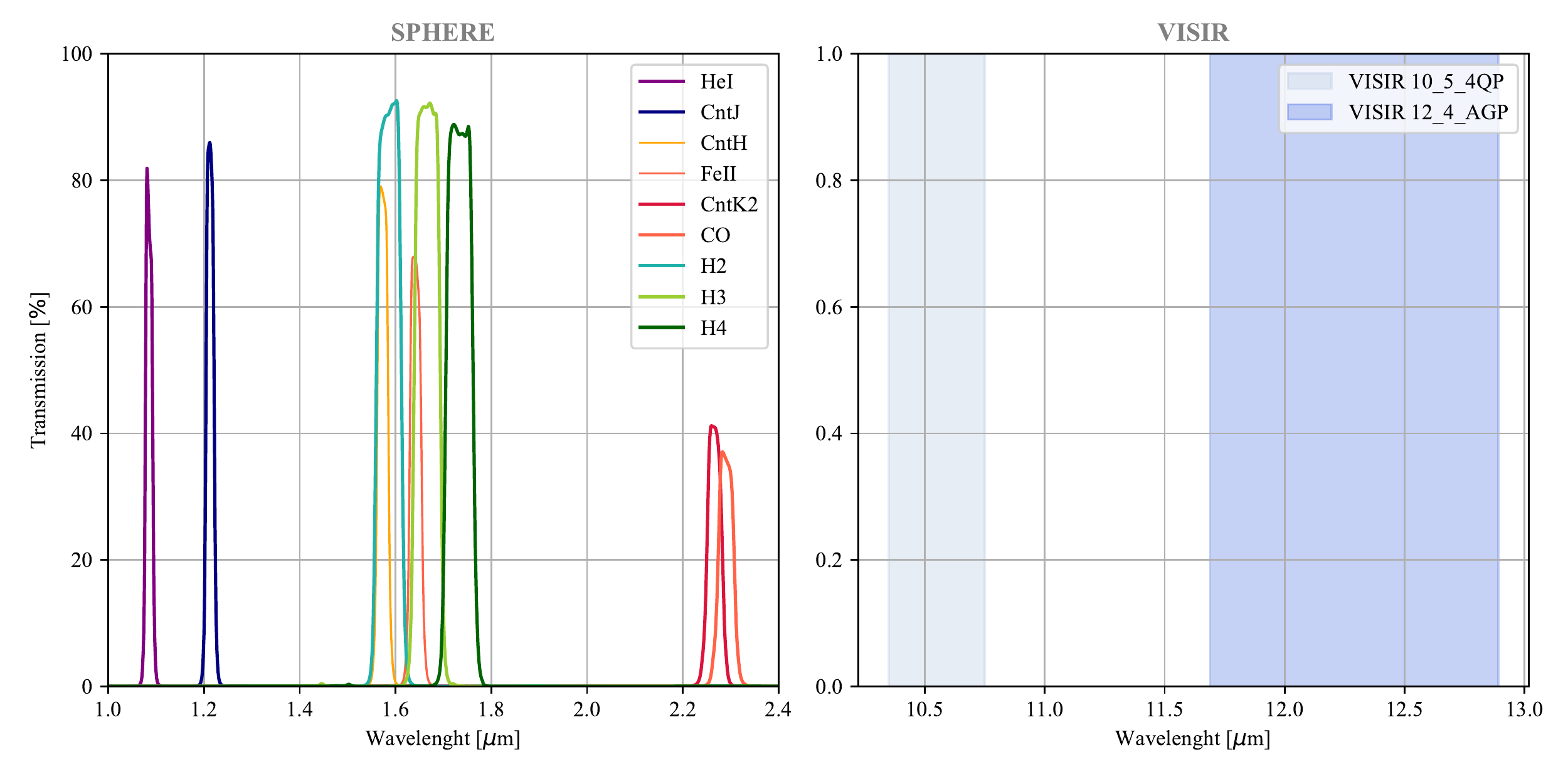} 
	\caption[example] 
	{\label{fig:filters}Transmission curves of all filters used in the present work. The exact transmission curves of the newly installed coronagraphic filters of VISIR are not yet available, so we present the bandwidth communicated by the ESO.}	
\end{figure}

\section{Detailed Log of observations}

\begin{table}[htbp]
\caption{Log of the SPHERE/IRDIS observations of WR104 and two PSF calibrators (4Sgr and TYC 6295-803-1). DIT is Detector Integration Time, NDIT the number of frames per exposure, NEXP the number of exposures for one observation, and SR the estimated Strehl ratio achieved by the AO system in the H band.}
	\label{tab:log_irdisfull}
	\centering
	\renewcommand{\arraystretch}{1.3}
	\begin{tabular}{c c c c c c c c c c c}
		\hline
		\hline
		No & Star & MJD & Filter & DIT[s]\texttimes NDIT & 			Dither\texttimes NEXP & $\Sigma$DIT[s] & Seeing [''] & SR\\
		\hline
		1 & WR104 & 57712.278 & CntH, FeII & 0.837$\times$5 & 4$\times$4$\times$3 & 200.85 & 0.39 & 0.82\\
         & 4Sgr & 57512.299 & CntH, FeII & 0.837$\times$5 &   4$\times$4$\times$3 & 200.85 & 0.56 & 0.92\\
       \hline
		2& WR104 & 57524.125 & H2, H3, H4 & 0.837$\times$7 & 4$\times$4$\times$2 & 187.5 & 1.13 & 0.73\\
        & 4Sgr & 57524.144 & H2, H3, H4 & 0.837$\times$7 & 4$\times$4$\times$2 & 187.5 & 1.2 & 0.78\\
        \hline
		3 & WR104 & 57590.050 & HeI, CntJ, CntK2, CO & 0.837$\times$6 & 4$\times$4$\times$3 & 241.02 & 0.45 & 0.87\\
         & TYC 6295-803-1 & 57590.09 & CntJ, CntK2 & 0.837$\times$5 & 4$\times$4$\times$3 & 200.88 & 0.44 & 0.87\\
       \hline
	\end{tabular}
\end{table}

\begin{table}[htbp]
	\caption{Log of the SPHERE/IFS observations of WR104 and 4Sgr (PSF calibrators).}
	\label{tab:log_ifs_full}
	\centering
	\renewcommand{\arraystretch}{1.3}
	\begin{tabular}{c c c c c c c c c c c c}
		\hline
		\hline
		No & Star & MJD & Filter & R & DIT[s]\texttimes NDIT & NEXP & $\Sigma$DIT[s] & Seeing [''] & SR\\
		\hline
		1& WR104 & 57524.125 & (Y-J) & 50 & 8$\times$10 & 6 & 480 & 1.13 & 0.75\\
		
		2 & 4Sgr & 57524.144 & (Y-J) & 50 & 8$\times$10 & 6 & 480 & 1.2 & 0.81\\
	\end{tabular}
\end{table}

\newpage
\section{All SPHERE images}

\begin{figure}[htbp]
	\centering
		\includegraphics[width=0.9\textwidth]{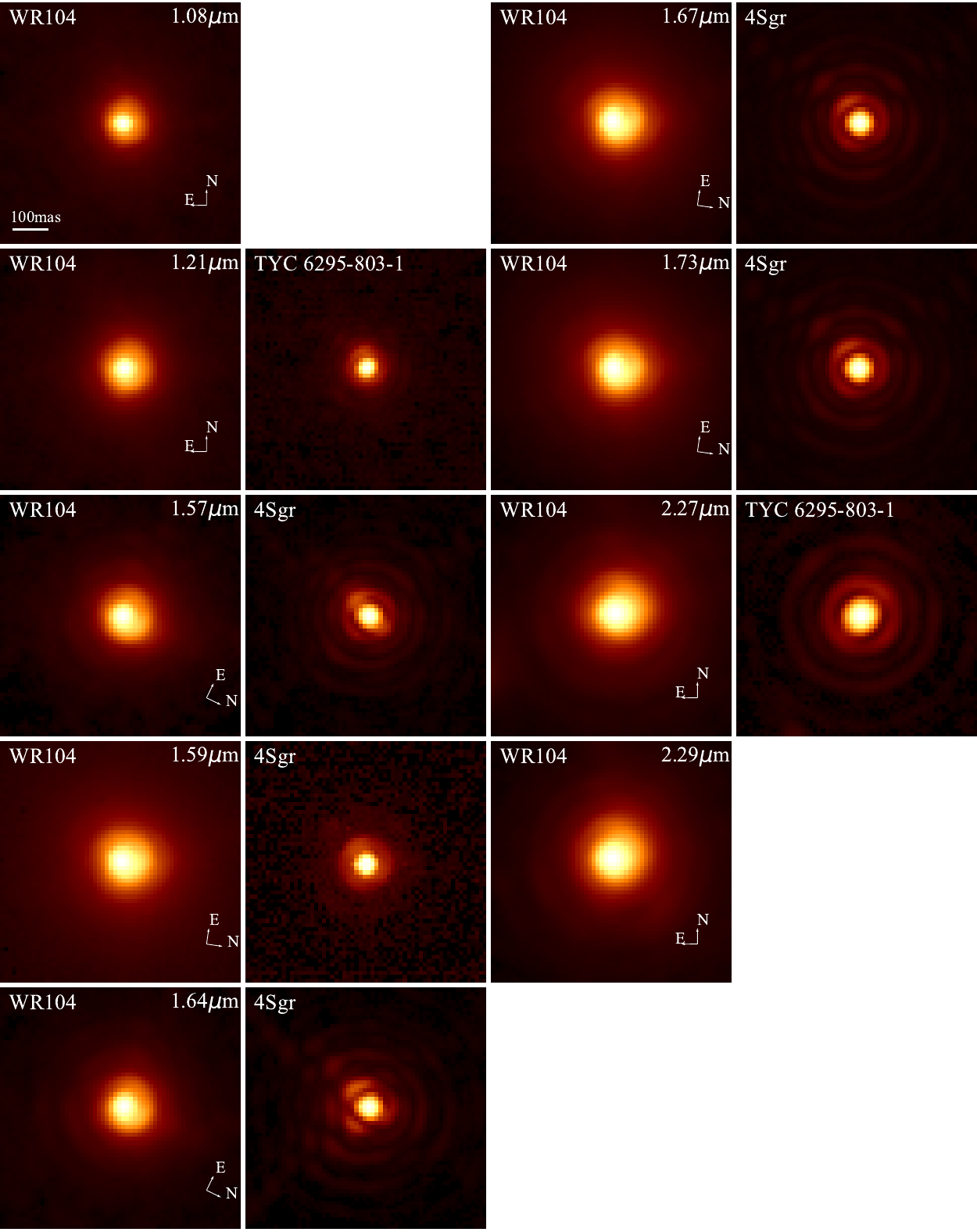} 
	\caption[] 
	{\label{fig:all_red}All reduced SPHERE images with the corresponding PSF calibrators. All images are rotated to be phased with the last SPHERE observation epoch (21 July 2016). The true orientation is shown on all panels.}
\end{figure}

\newpage
\begin{figure}[htbp]
		\centering
			\includegraphics[width=.8\textwidth]{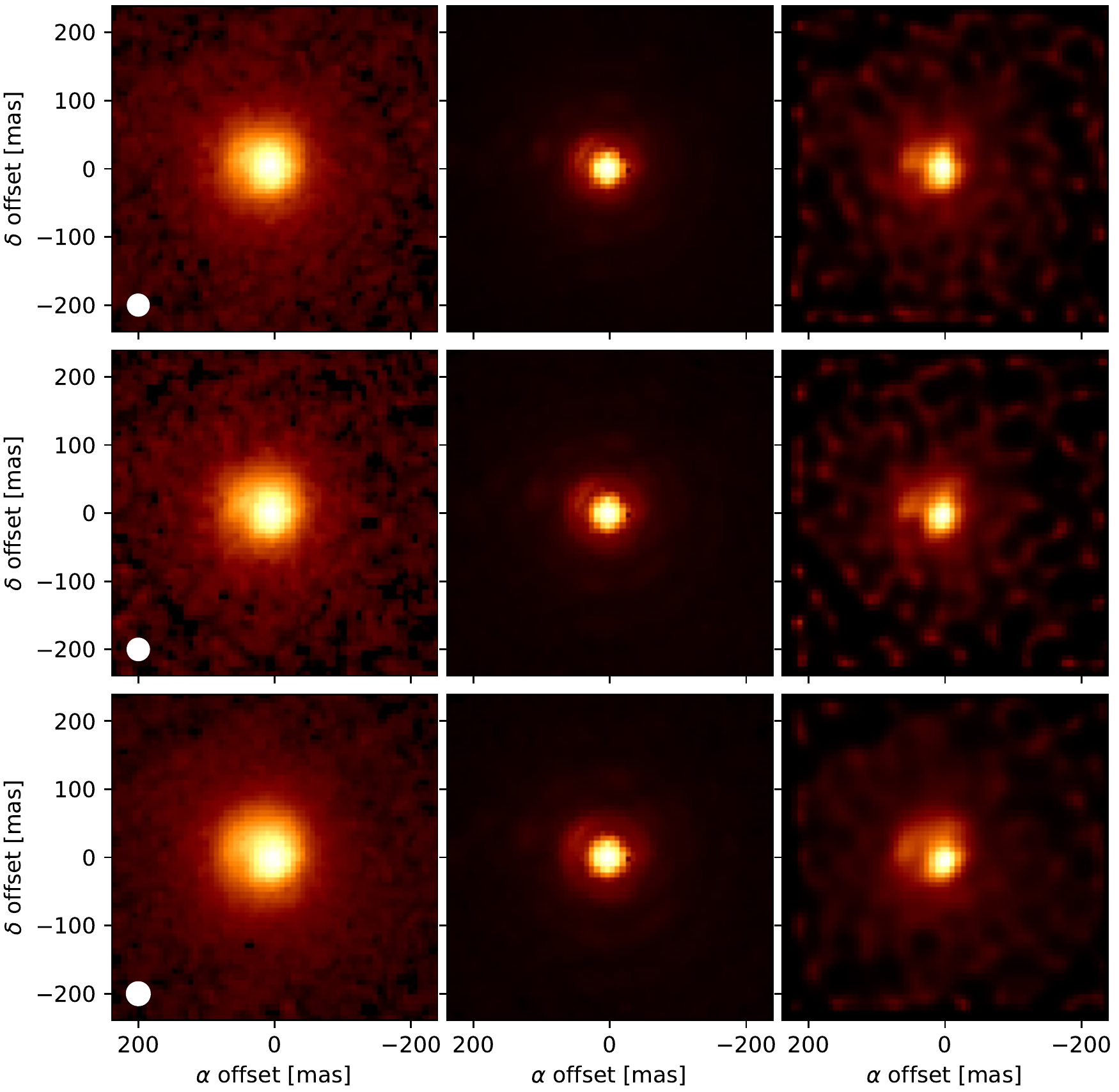} 
		\caption[example] 
		{ \label{fig:IFS_data} IFS images in the different part of the spectra (see Fig.~\ref{fig:spectrum}). \textbf{Top: }blue part of Fig.~\ref{fig:spectrum} (Y band, $\lambda=1.01\pm0.05\mu m$). \textbf{Middle: } green part of Fig.~\ref{fig:spectrum} (He line, $\lambda=1.09\pm0.03\mu m$). \textbf{Bottom: } purple part of Fig.~\ref{fig:spectrum} (J band, $\lambda=1.20\pm0.07\mu m$). We represent the reduced image (left), the PSF calibrator (middle) and the deconvolution (right). The FWHM of the PSF is also represented by a white circle in all the panels of the left column.}
\end{figure}

\newpage
\section{Radial profiles and uncertainties}

\begin{figure}[h]
\centering
		\includegraphics[width=1\textwidth]{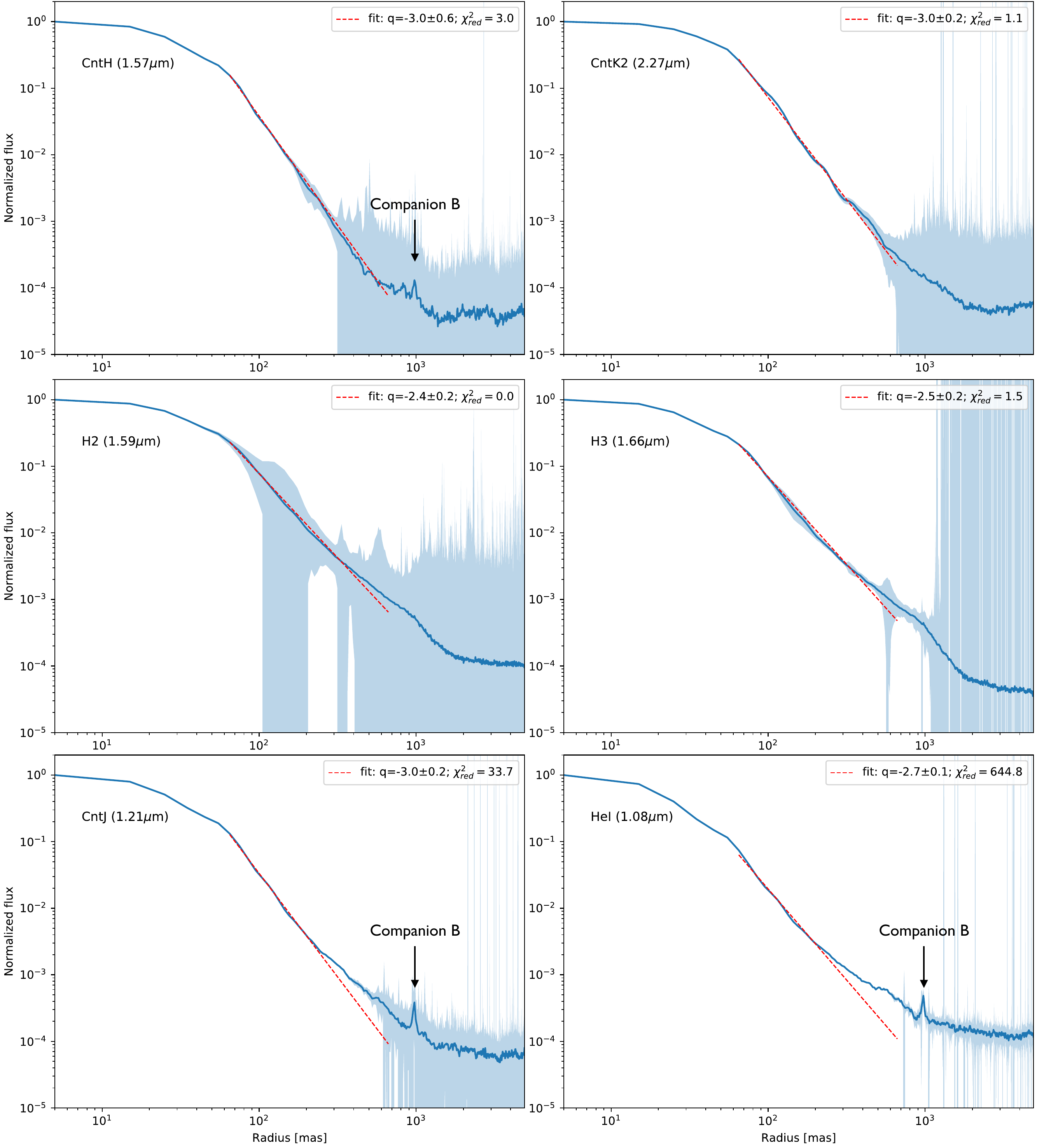} 
	\caption{Results of a power-law fit on the radial profiles. The top panels represent two good fits with a representative estimation of the uncertainty in the data (in light blue). The middle panels show a good fit but with data of poor quality. The bottom panels show the deviation of the radial profile from a power-law in the J band.\label{fig:rad_prf_quality}}	
\end{figure}

\twocolumn
\section{Complementary model fitting results}

\begin{figure}[h!]
\centering
		\includegraphics[width=.49\textwidth]{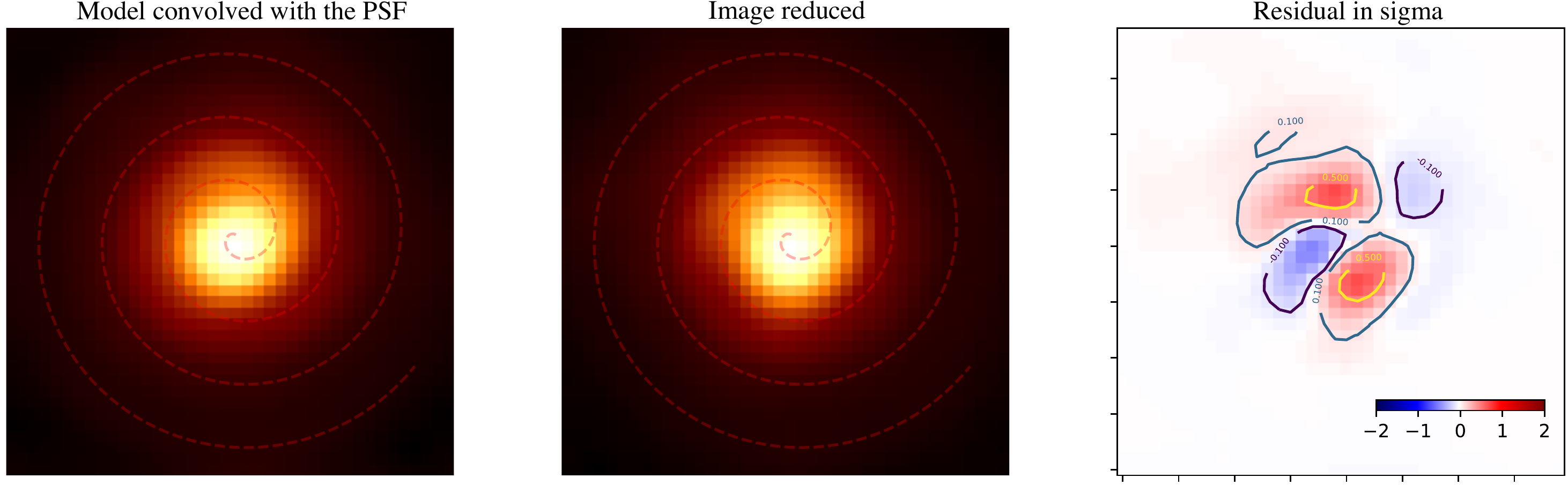} 
	\caption{Comparison of the best fit model in the K band and the reduced image. The residuals are represented in terms of $\sigma$ (standard deviation of the image). \label{fig:bestfit_compa}}	
\end{figure}

In addition to Section~\ref{simplemodel}, we present a comparison of the best fit phenomenological model with the K band image. The step and orientation of the spiral is represented in Fig.~\ref{fig:bestfit_compa} as a dashed red line. The K band data are well-fitted by the model, with residuals within $1\sigma$ compared to the standard deviation.

\begin{figure}[h]
	\centering
		\includegraphics[width=.48\textwidth]{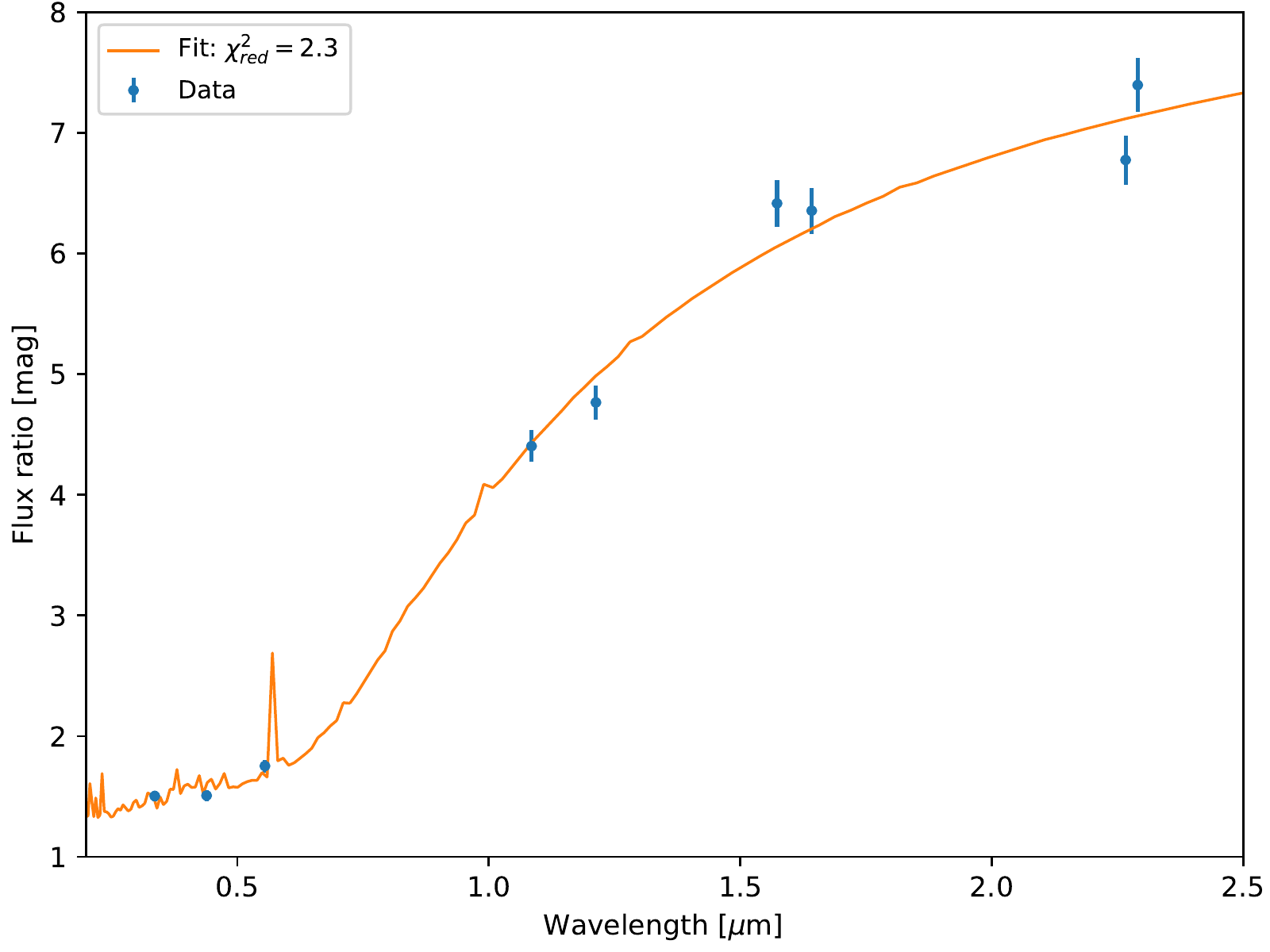} %
	\caption{\label{fig:rflux} 
    Flux ratio between the central binary star surrounded by dust and the companion B star. Our best fit is represented in orange ($\rm T_{dust}=2200\;K$, $\rm T_{B}=45,000\;K$).}
\end{figure}

\begin{figure}[h]
	\centering
		\includegraphics[width=.48\textwidth]{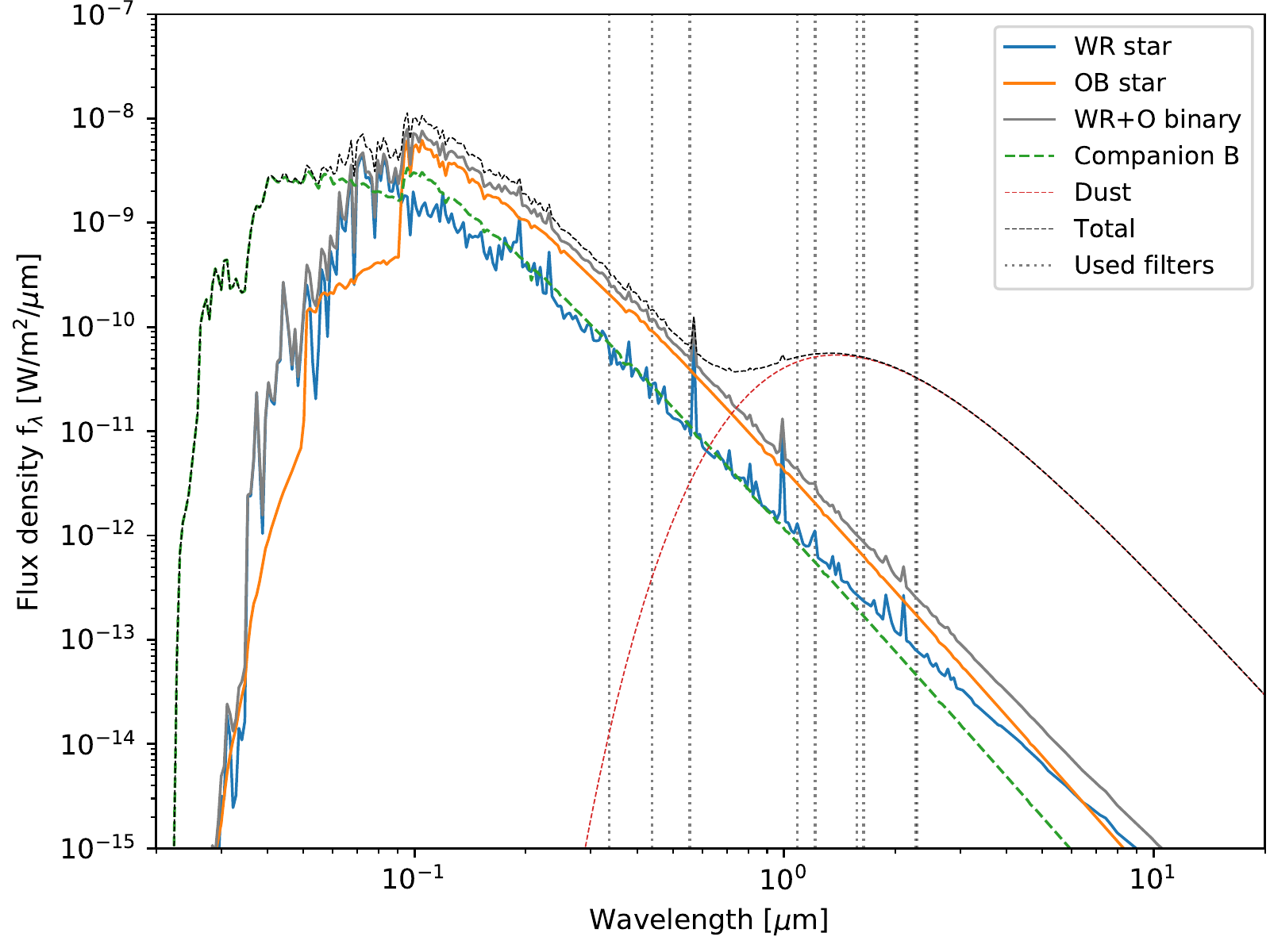} %
	\caption{\label{fig:sed_compB} 
   Best fits of the flux density from: 1) two Kurucz atmosphere models representing the companion B (green dashed line) and the OB-type inner companion (orange line), 2) one PoWR model representing the WC9 star (blue line), and 3) one blackbody model representing the dusty pinwheel (red dashed line). The flux ratio is obtained by comparison between the WR+OB flux density (grey line) and the companion B. We also show the total flux density (black dashed line) and position of the different filters used (vertical black dotted lines).}
\end{figure}

\medskip
As presented in Section \ref{compB_study}, we used different atmosphere models to determine the effective temperature of the third component of WR104, i.e.\ companion B. Fig.~\ref{fig:rflux} shows the results of the best fit of the flux ratio between the central binary star and the companion star. We present the resulting flux density at the distance of WR104 ($\rm D=2.58\pm0.12$\,kpc, see Sec.~\ref{distance}) of the four components in Fig.~\ref{fig:sed_compB}. The total flux density is comprised of two Kurucz atmosphere models representing the companion B and the OB-type inner companion: a Potsdam Wolf-Rayet (PoWR) model representing the WR star and a blackbody representing the dust around the central binary.

The infrared excess of WR104 compared to the companion indicates the presence of relatively hot dust around the central binary star, i.e.\ the pinwheel nebula.  Given the relative flatness of the flux ratio in the B and V bands, the HST data provide important information about the stellar temperature. This indicates that companion B star has an effective temperature comparable to the OB star of the inner binary (which dominates the total flux), and is probably an OB type star itself, thereby confirming the findings of  \citetads{2002ASPC..260..407W}.

\end{document}